%% file: draft1.tex
\documentclass[prd,aps,a4paper,superscriptaddress,twocolumn,nofootinbib]{revtex4}
\usepackage{graphicx}
\usepackage{color}
\usepackage{xcolor}
\usepackage{dcolumn}
\usepackage{bm}
\usepackage{slashed}
\usepackage{amsmath}
\usepackage{latexsym}
\usepackage{amssymb}
\usepackage{mathrsfs}
\usepackage{amsfonts}
\usepackage{longtable}
\usepackage{physics}
\usepackage{xspace}
\allowdisplaybreaks

\begin{document}
\title{Effective-One-Body Numerical-Relativity waveform model for Eccentric spin-precessing binary black hole coalescence}

\author{Xiaolin Liu}
\affiliation{Institute for Frontiers in Astronomy and Astrophysics, Beijing Normal University, Beijing 102206, China}
\affiliation{Department of Astronomy, Beijing Normal University,
Beijing 100875, China}
\author{Zhoujian Cao
\footnote{corresponding author}} \email[Zhoujian Cao: ]{zjcao@amt.ac.cn}
\affiliation{Institute for Frontiers in Astronomy and Astrophysics, Beijing Normal University, Beijing 102206, China}
\affiliation{Department of Astronomy, Beijing Normal University,
Beijing 100875, China}
\affiliation{School of Fundamental Physics and Mathematical Sciences, Hangzhou Institute for Advanced Study, UCAS, Hangzhou 310024, China}
\author{Zong-Hong Zhu}
\affiliation{Institute for Frontiers in Astronomy and Astrophysics, Beijing Normal University, Beijing 102206, China}
\affiliation{Department of Astronomy, Beijing Normal University,
Beijing 100875, China}

\begin{abstract}
Waveform models are important to gravitational wave data analysis. People recently pay much attention to the waveform model construction for eccentric binary black hole coalescence. Several Effective-One-Body Numerical-Relativity waveform models of eccentric binary black hole coalescence have been constructed. But none of them can treat orbit eccentricity and spin-precessing simultaneously. The current paper focuses on this problem. The authors previously have constructed waveform model for spin-aligned eccentric binary black hole coalescence $\texttt{SEOBNRE}$. Here we extend such waveform model to describe eccentric spin-precessing binary black hole coalescence. We calculate the 2PN orbital radiation-reaction forces and the instantaneous part of the decomposed waveform for a general spinning precessing binary black hole system in effective-one-body (EOB) coordinates.
We implement these results based on our previous $\texttt{SEOBNRE}$ waveform model. We have also compared our model waveforms to both SXS and RIT numerical relativity waveforms. We find good consistency between our model and numerical relativity. Based on our new waveform model, we analyze the impact of the non-perpendicular spin contributions on waveform accuracy. We find that the non-perpendicular spin contributions primarily affect the phase of the gravitational waveforms. For the current gravitational wave detectors, this contribution is not significant. The future detectors may be affected by such non-perpendicular spin contributions. More importantly our $\texttt{SEOBNRE}$ waveform model, as the first theoretical waveform model to describe eccentric spin-precessing binary black hole coalescence, can help people to analyze orbit eccentricity and spin precession simultaneously for gravitational wave detection data.
\end{abstract}

\maketitle

\section{Introduction}
After several observation runs, the LIGO and Virgo \cite{advLIGO,advVirgo} ground-based gravitational wave detectors have detected a large number of binary black hole (BBH) \cite{PhysRevLett.116.061102,GWTC3}, binary neutron stars (BNS) \cite{PhysRevLett.119.161101} and binary black hole-neutron star (BH-NS) \cite{BH_NS} events. Along with the increasing detector sensitivity, we expect that in future observations, LIGO-Virgo-Kagra \cite{KAGRA}, as well as the under-construction ET \cite{ET}, CE \cite{CE}, and space-based detectors such as LISA \cite{LISA}, Taiji \cite{Taiji}, Tianqin \cite{Tianqin}, will be able to detect more dynamic features of binary black hole systems.
This will provide us valuable physics information about extremely strong gravitational field and provide powerful support for the study of gravitational theory, galaxy formation theory, and cosmology, as well as the verification of general relativity.

Merger systems of compact binary object systems are the most common gravitational wave sources in the universe. Their dynamics and related gravitational waves have been studied for many years \cite{Einstein_1938,Lorentz_1937,Ohta_1973,Damour_1985,Blanchet_2006}. The research on this yopic is still ongoing \cite{Jaranowski_2012,Damour_2016,Bernard_2018}.
The detection and data analysis of gravitational waves depend on the construction of gravitational waveform templates \cite{PhysRevD.46.5236,PhysRevD.49.1723,PhysRevD.58.063001}.
Many years ago, the designed probing targets of LIGO are gravitational wave events from compact binary star systems in the merger phase.
Near the merger stage, people expect that the eccentricity of the binary system has been reduced to almost zero.
Therefore, many commonly used templates now are based on circular orbit assumption for two-body systems.
In recent years, people have begun to pay attention to gravitational waves from two-body systems along eccentric orbit, and have found some traces of eccentricity residuals in LIGO-Virgo data \cite{PhysRevLett.125.101102}.
This situation calls for more accurate and comprehensive gravitational wave templates for general two-body systems.

So far, many eccentric models without spin precession and spin-precession circular models have been established.
Among them are templates based on the effective one-body model, such as the $\texttt{TEOBResumS}$ \cite{PhysRevD.101.101501,PhysRevD.103.104021} model, the $\texttt{SEOBNRE}$ \cite{PhysRevD.96.044028,PhysRevD.101.044049_validSEOBNRE,PhysRevD.103.124053,Liu_2022,2023IJMPD..3250015L} model and $\texttt{SEOBNRv4E}$ \cite{PhysRevD.105.044035,PhysRevD.104.024046} model.
These models assume that the spin direction of the binary black hole is parallel to the orbital angular momentum, in which case the direction of the orbital angular momentum remains unchanged.
If the spin direction of one black hole is not parallel to the direction of the orbital angular momentum, the orbital direction of the binary black hole system will change due to the interaction between the spin of the black hole and the orbital angular momentum, which is called spin precession \cite{PhysRevD.49.6274}.
Currently, many models have considered the spin-precession binary black hole system on circular orbits, such as the IMRPhenomP \cite{PhysRevD.103.104056} model and the $\texttt{SEOBNRv4P}$ \cite{PhysRevD.89.084006,PhysRevD.102.044055} model.
Recently, the next generation spin precession framework $\texttt{SEOBNRv5}$ with higher accuracy and computational efficiency has been published \cite{seobnrv5phm,seobnrv5}.

Besides the afore mentioned complete inspiral-merger-ringdown waveform models, there are other phenomenological and post-Newtonian (PN) approximated waveform models for eccentric BBH. In \cite{PhysRevD.97.024031_ENIGMA,PhysRevD.100.044016_ENIGMA_CalibNR,PhysRevD.103.084018} Eccentric, Nonspinning, Inspiral-Gaussian-process Merger Approximant $(\texttt{ENIGMA})$ model was proposed, which combines analytical and NR results using machine learning algorithms. The authors of \cite{PhysRevD.98.044015} combine PN inspiral waveform for eccentric BBH and circular NR waveform for merger and ringdown. The surrogate eccentric waveform model based on numerical relativity simulations was proposed in \cite{PhysRevD.103.064022} for spinless black holes. The authors of \cite{PhysRevD.103.124011,PhysRevD.107.124061} constructed an eccentric waveform model for spinless black holes through fitting the difference between the circular waveform and eccentric waveform. Pioneered by Peters \cite{PhysRev.136.B1224_Peters}, post-Newtonian eccentric waveform models \cite{PhysRevD.91.084040_3PNWFInst,PhysRevD.93.064031,PhysRevD.93.124061,PhysRevD.100.044018_3PNWFtail,PhysRevD.100.084043_3PNWFmem,PhysRevD.102.084042,2017CQGra..34d4003L,2017CQGra..34m5011L,2018CQGra..35w5006M,2020CQGra..37g5008L,2021CQGra..38a5005L} include post-circular (PC) waveform model \cite{PhysRevD.80.084001_PCmode}, enhanced post-circular (EPC) waveform model \cite{PhysRevD.90.084016_EPCmodel}, x-model \cite{PhysRevD.82.024033_xmodel} and others.

In order to aid the gravitational wave data analysis, we need complete and accurate waveform models for generic binary systems. Especially waveform models of eccentric spin-precessing binary black hole coalescence is needed \cite{2020ApJ...903L...5R,2023MNRAS.519.5352R,2023NatAs...7...11G,2022NatAs...6..344G}. In the current paper we extend $\texttt{SEOBNRE}$ to construct such a waveform model. Our waveform model is based on effective one body (EOB) numerical relativity framework. The framework has been introduced in \cite{PhysRevD.96.044028}. As the two major blocks of the framework, we describe the extended EOB dynamics in the next section and the extended waveform formula in Sec.~\ref{sec3}. After that we check the performance of our waveform model by comparing the model waveform to numerical relativity waveform in Sec.~\ref{sec4}. Finally we conclude the paper with some discussions in the last section.

Throughout the paper we use geometric unit system with the speed of light and gravitational constant taking one ($c=G=1$). Some equations show $c$ explicitly for the clarity of order expansion.

\section{Effective-one-body dynamics for eccentric spin-pressing BBH}
As usual we use Hamiltonian to describe the EOB dynamics. The EOB Hamiltonian of a BBH system with mass $m_{1,2}$ reads \cite{PhysRevD.59.084006}
\begin{align}
&H_{\text{EOB}} = M\sqrt{1+2\nu \left( \hat{H}_{\text{eff}}-1 \right)}, \label{EOBE}\\
&\hat{H}_{\text{eff}}\equiv\frac{H_{\text{eff}}}{M\nu},
\end{align}
where the total mass $M=m_1+m_2$ and symmetric mass ratio $\nu=m_1m_2/M^2$. The effective Hamiltonian $H_{\text{eff}}$ depends on BH's spin $\pmb{S}_{1,2}:=m_{1,2}^2\pmb{\chi}_{1,2}$, where $\pmb{\chi}$ is dimensionless spin vector satisfying $-1<|\pmb{\chi}|<1$. We borrow the function form of $H_{\text{eff}}$ from \cite{PhysRevD.81.084024,PhysRevD.84.104027}. Then the dynamics of eccentric spin-pressing BBHs can be described by the following canonical equations
\begin{align}
&\dv{\pmb{r}}{t}=\pdv{H_{\text{EOB}}}{\pmb{p}}, \\
&\dv{\pmb{p}}{t}=-\pdv{H_{\text{EOB}}}{\pmb{r}}-\pmb{F}, \\
&\dv{\pmb{S}_{1,2}}{t}=\pdv{H_{\text{EOB}}}{\pmb{S}_{1,2}}\times\pmb{S}_{1,2}+\pmb{F}^s_{1,2},
\end{align}
where $\pmb{r}$ is relative position vector, and $\pmb{F}, \pmb{F}^s_{1,2}$ are radiation reaction forces for momentum $\pmb{p}$ and BH spins respectively.

The spin radiation reaction forces $\pmb{F}^s_{1,2}$ can be written as \cite{PhysRevD.96.084064,PhysRevD.96.084065}
\begin{widetext}
\begin{align}
&\pmb{F}_{1}^s=\frac{4 p_r}{15r^4}(\pmb{L}\times\pmb{S}_1)\left[\nu\left(-22\frac{1}{r}+36p^2-60p_r^2\right)+\frac{1}{2}(1-2\nu-\delta)\left(16\frac{1}{r}-48p^2+75p_r^2\right)\right] \nonumber\\
&+\frac{2\nu}{15r^6}\left[6r^2p_r(\pmb{S}_1\times\pmb{S}_2)\left(11\frac{1}{r}-18p^2+30p_r^2\right)-3r(\pmb{S}_1\times\pmb{p})\left(\pmb{r}\cdot\pmb{S}_2-\frac{1-\delta-2\nu}{2\nu}\pmb{r}\cdot\pmb{S}_1\right)\left(8\frac{1}{r}-9p^2+45p_r^2\right) \right. \nonumber\\
&\left. - (\pmb{r}\times\pmb{S}_1)\left[15p_r\left(\pmb{r}\cdot\pmb{S}_2-\frac{1-\delta-2\nu}{2\nu}\pmb{r}\cdot\pmb{S}_1\right)\left(2\frac{1}{r}-3p^2+7p_r^2\right)+2r\left(\pmb{p}\cdot\pmb{S}_2-\frac{1-\delta-2\nu}{2\nu}\pmb{p}\cdot\pmb{S}_1\right)\left(\frac{8}{r}-9p^2+45p_r^2\right)\right]\right] \nonumber\\
& - \frac{2(1-\delta)}{5r^5}\left[4\left(\frac{1}{r}-3p^2+15p_r^2\right)\left[(\pmb{r}\times\pmb{S}_1)\pmb{r}\cdot\pmb{S}_1+(\pmb{p}\times\pmb{S}_1)\pmb{r}\cdot\pmb{S}_1\right]+15\frac{1}{r}p_r(\pmb{r}\times\pmb{S}_1)(\pmb{r}\cdot\pmb{S}_1)(3p^2-7p_r^2)\right. \nonumber\\
&\left.-30rp_r(\pmb{p}\times\pmb{S}_1)(\pmb{p}\cdot\pmb{S}_1)\right],
\end{align}
\end{widetext}
where $\pmb{L}=\pmb{r}\times\pmb{p}$ is the orbital angular momentum, $p_r=\pmb{r}\cdot\pmb{p}/r$, and $\delta=(m_1-m_2)/M,\pmb{F}_{2}^s=\pmb{F}_{1}^s(1\leftrightarrow 2)$. In \cite{PhysRevD.96.084064,PhysRevD.96.084065}, Harmonic coordinate is used. Since the Harmonic coordinates equals to the EOB coordinates to leading PN order, we apply the above formula of the spin radiation reaction forces $\pmb{F}^s_{1,2}$ to EOB coordinate directly. In the
quasi-circular limit $p_r=0$, the above equations indicate that $\pmb{F}^s_{\text{qc},1,2}\approx0$ consequently \cite{seobnrv5}.

For general eccentric spin-precessing cases, the radiation reaction force satisfies a balance equation with an extra Schott terms $\dot{E}_{\text{Schott}}$ and $\dot{\pmb{J}}_{\text{Schott}}$ \cite{PhysRevD.86.124012,PhysRevD.104.024046}
\begin{align}
\dot{E}_{\text{sys}}+\dot{E}_{\text{Schott}}=&\mathcal{F}, \nonumber\\
\dot{\pmb{J}}_{\text{sys}}+\dot{\pmb{J}}_{\text{Schott}}=&\pmb{\mathcal{J}}, \label{eq_balance}
\end{align}
where $E_{\text{sys}},\pmb{J}_{\text{sys}}$ are the energy and total angular momentum of the binary system respectively, $\mathcal{F}=\dd{E_{\text{gw}}}/\dd{t}$ and
$\pmb{\mathcal{J}}$ respectively represent the energy flux and the angular momentum flux of gravitational waves at infinity. The post-Newtonian approximation of the energy flux and the angular momentum flux of gravitational waves has been calculated in \cite{PhysRevD.47.R4183,PhysRevD.87.044009}. For our usage we transform the Harmonic coordinate in these calculation results to EOB coordinate. We follow the detail transformation given in \cite{PhysRevD.104.024046}.

The radiation reaction forces act on the binary system and cause the energy and angular momentum loss \cite{PhysRevD.75.064017}
\begin{align}
&\dot{E}_{\text{sys}}=\dot{\pmb{r}}\cdot \pmb{F}, \nonumber\\
&\dot{\pmb{J}}_{\text{sys}}=\pmb{r}\times\pmb{F}+\pmb{F}^s_{1}+\pmb{F}^s_{2} \label{eq_rrforce}.
\end{align}

Substituting Eq.~(\ref{eq_rrforce}) into Eq.~(\ref{eq_balance}), we can solve the radiation reaction force and get
\begin{align}
&\pmb{F}=\frac{1}{\dot{\pmb{r}}\cdot\pmb{r}}\left[\pmb{r}(\mathcal{F}-\dot{E}_{\text{Sch}})-\dot{\pmb{r}}\times\left[\pmb{\mathcal{J}}-\dot{\pmb{J}}_{\text{Sch}}-(\pmb{F}^s_{1}+\pmb{F}^s_{2})\right]\right].\label{eq1}
\end{align}

Regarding to the Schott term $E_{\text{Sch}}$, we borrow the instant term shown in the first equation of Eq.~(31) of \cite{PhysRevD.104.024046}. And more we borrow the SO term shown in the first equation of Eq.~(28) of \cite{PhysRevD.104.024046} and replace $\chi_{1,2}$ with $\pmb{L}\cdot\pmb{S}_{1,2}$. For the SS term we adopt form
\begin{widetext}
\begin{align}
&\qquad\qquad E^{\text{SS}}_{\text{Sch}}=\frac{\eta^2p_r}{r^4c^4}\left[\pmb{S}_1^2(a_1p_r^2+a_2p^2+\frac{a_3}{r})+ (\pmb{S}_1\cdot\pmb{S}_2)(a_4p_r^2+a_5p^2+\frac{a_6}{r})+\pmb{S}_2^2(a_7p_r^2+a_8p^2+\frac{a_9}{r})\right]+\nonumber \\
&\frac{1}{r^7c^4}\left[(\pmb{p}\cdot\pmb{S}_1)(\pmb{r}\cdot\pmb{S}_1)\left(\frac{h_1}{r}+h_2 p^2+h_3p_r^2\right)+(\pmb{r}\cdot\pmb{S}_1)(\pmb{p}\cdot\pmb{S}_2)\left(\frac{h_4}{r}+h_5 p^2+h_6p_r^2\right)+1\leftrightarrow 2\right] + \nonumber\\
&\frac{p_r}{r^7c^4}\left[d_1(\pmb{r}\cdot\pmb{S}_1)^2+d_2(\pmb{r}\cdot\pmb{S}_1)(\pmb{r}\cdot\pmb{S}_2)+1\leftrightarrow 2\right],\label{ESchSS}
\end{align}
\end{widetext}

Regarding to the Schott term $\pmb{J}_{\text{Sch}}$, we borrow the instant term for $\pmb{J}_{\text{Sch}}$ shown in the second equation of Eq.~(31) of \cite{PhysRevD.104.024046} and replace the magnitude of the angular momentum $L$. For the SO term, we need to introduce components besides the one along the angular momentum direction $\pmb{L}$
\begin{align}
&\pmb{J}^{\text{SO}}_{\text{Sch}}=\frac{\eta^2p_r}{r^2c^3}\left[\frac{\pmb{L}}{r}\left(\pmb{L}\cdot\pmb{S}_1(\beta_1p_r^2+\beta_2p^2+\frac{\beta_3}{r})\right) + \right.\nonumber\\
&\left.\pmb{p}\left(\frac{\beta_4}{r}\pmb{r}\cdot\pmb{S}_1+\beta_5\pmb{p}\cdot\pmb{S}_1\right)+\pmb{S}_1(\beta_6p_r^2+\beta_7p^2+\frac{\beta_8}{r})\right] + \nonumber\\
&\frac{\beta_9\eta^2\pmb{r}}{r^4c^3}\pmb{p}\cdot\pmb{S}_1+1\leftrightarrow 2.
\end{align}
For the SS term we adopt form
\begin{widetext}
\begin{align}
&\qquad\qquad\pmb{J}=\frac{\eta^2p_r\pmb{L}}{r^4c^4}\left[a_{10}\pmb{S}_1^2+a_{11}\pmb{S}_1\cdot\pmb{S}_2+a_{12}\pmb{S}_2^2+\frac{1}{r^2}\left[f_1(\pmb{r}\cdot\pmb{S}_1)(\pmb{r}\cdot\pmb{S}_1)+f_2(\pmb{r}\cdot\pmb{S}_1)(\pmb{r}\cdot\pmb{S}_2)+f_3(\pmb{r}\cdot\pmb{S}_2)(\pmb{r}\cdot\pmb{S}_2)\right]+\right.\nonumber\\
&\left.r\left[f_4(\pmb{p}\cdot\pmb{S}_1)(\pmb{p}\cdot\pmb{S}_1)+f_5(\pmb{p}\cdot\pmb{S}_1)(\pmb{p}\cdot\pmb{S}_2)+f_6(\pmb{p}\cdot\pmb{S}_2)(\pmb{p}\cdot\pmb{S}_2)\right]\right]+\frac{\eta^2p_r}{r^4c^4}\left[\pmb{p}\times\pmb{S}_1\left(b_1(\pmb{r}\cdot\pmb{S}_1)+b_2(\pmb{r}\cdot\pmb{S}_2)\right)+\frac{\pmb{r}\times\pmb{S}_1}{r}\right.\nonumber\\
&\left.\left(\left(\frac{b_3}{r}+b_4p_r^2+b_5p^2\right)(\pmb{p}\cdot\pmb{S}_1)+\left(\frac{b_6}{r}+b_7p_r^2+b_8p^2\right)(\pmb{p}\cdot\pmb{S}_2)\right)+1\leftrightarrow 2\right]+\frac{\eta^2}{r^2c^4}\left[\pmb{p}\times\pmb{S}_1\left((\pmb{p}\cdot\pmb{S}_1)\left(c_1p^2+c_2p_r^2+\frac{c_3}{r}\right)+ \right.\right.\nonumber \\
&\left.\left.(\pmb{p}\cdot\pmb{S}_2)\left(c_4p^2+c_5p_r^2+\frac{c_6}{r}\right)\right)+\frac{\pmb{r}\times\pmb{S}_1}{r^3}\left((\pmb{r}\cdot\pmb{S}_1)\left(d_1p^2+d_2p_r^2+\frac{d_3}{r}\right)+(\pmb{r}\cdot\pmb{S}_2)\left(d_4p^2+d_5p_r^2+\frac{d_6}{r}\right)\right)+1\leftrightarrow 2\right]+\nonumber\\
&\frac{\eta^2\pmb{L}}{r^6}\left[(\pmb{r}\cdot\pmb{S}_1)(\pmb{p}\cdot\pmb{S}_1)\left(\frac{e_1}{r}+e_2p^2+e_3p_r^2\right)+(\pmb{r}\cdot\pmb{S}_1)(\pmb{p}\cdot\pmb{S}_2)\left(\frac{e_4}{r}+e_5p^2+e_6p_r^2\right)+1\leftrightarrow 2\right]
.\label{JSchSS}
\end{align}
\end{widetext}

In the above equations about Schott terms we have introduced a set of to-be-determined coefficients. In order to fix them, we tried condition
\begin{align}
\frac{\pmb{F}\cdot\pmb{r}L}{|\pmb{r}\times\pmb{F}|rp_r}=&1+\order{p_r^2}\label{EquatorialCircCondition}
\end{align}
used in \cite{PhysRevD.89.084006}. But we find this condition is not enough to determine the set of coefficients mentioned above.

In fact the condition (\ref{EquatorialCircCondition}) can be looked as a nearly circular approximation. Keeping this fact in mind and we note that the radiation reaction force in circular limit should take the form \cite{PhysRevD.89.084006}
\begin{align}
\pmb{F}|_{p_r=0,\dot{p}_r=0}=\frac{\mathcal{F}_{\text{qc}}}{\Omega_{\text{qc}}|\pmb{L}|}\pmb{p},
\end{align}
where $\Omega_{\text{qc}}$ means the orbital angular velocity. The above equation implies
\begin{align}
&\frac{\Omega_{\text{qc}} L}{\mathcal{F}_{\text{qc}}}\pmb{F}=\pmb{p}\left[\order{1}+\order{p_r}+\order{\dot{p}_r}\right]+\nonumber\\
&\pmb{r}\left[\order{p_r}+\order{\dot{p}_r}\right]+\pmb{L}\left[\order{p_r}+\order{\dot{p}_r}\right] + \nonumber\\
&\pmb{S}_1\left[\order{p_r}+\order{\dot{p}_r}\right]+\pmb{S}_2\left[\order{p_r}+\order{\dot{p}_r}\right].\label{PrecCircCondition}
\end{align}

We plug the Schott terms into (\ref{eq1}) and expand the vector $\pmb{F}$ according to the form shown in the above equation. The condition (\ref{PrecCircCondition}) tells us that in the $\pmb{p}$ direction, the lowest order in the expansions using $p_r$ and $\dot{p}_r$ is $\order{1}$, while in other directions, such as $\pmb{r}$ and $\pmb{L}$, the lowest order is at least $\order{p_r}$ or $\order{\dot{p}_r}$. These facts give us a set of equations. We solve these equations and fix the to-be-determined coefficients introduced in the Schott terms. After that combining the energy flux, angular momentum flux, Schott terms and $\pmb{F}^s_{1,2}$ we archive the form of the radiation reaction force
\begin{align}
&\pmb{F}= \left(f^r_{\text{ns}} + \frac{1}{c^3}f^r_{\text{so}} + \frac{1}{c^4}f^r_{\text{ss}}\right)\pmb{r} + \left( f^p_{\text{ns}} + \frac{1}{c^3}f^p_{\text{so}} + \frac{1}{c^4}f^p_{\text{ss}} \right)\pmb{p} \nonumber\\
& + \frac{1}{c^3}f^L_{\text{so}}\pmb{L} + \frac{1}{c^4}f^1_{\text{ss}}\pmb{S}_1 + \frac{1}{c^4}f^2_{\text{ss}}\pmb{S}_2,
\end{align}
The coefficients $f$s are listed in the Appendix~\ref{appA}.

Next, we re-sum this radiation reaction force into a factorized form,
\begin{align}
&\pmb{F}=\frac{\mathcal{F}}{\Omega L}(\mathcal{F}^e_p\pmb{p}+\mathcal{F}^e_r\pmb{r}+\mathcal{F}^e_L\pmb{L}+\mathcal{F}^e_1\pmb{S}_1+\mathcal{F}^e_2\pmb{S}_2),\label{eq2}\\
&\Omega\equiv|\pmb{r}\times\dot{\pmb{r}}|/r^2,\\
&L\equiv|\pmb{r}\times\pmb{p}|.
\end{align}
The factors $\mathcal{F}^e_{...}$ are listed in the supplementary materiel. Eq.~(\ref{eq2}) is used in the updated $\texttt{SEOBNRE}$ code to calculate the radiation reaction force. Numerically we calculate $\mathcal{F}_{\text{qc}}$ with the waveform got in the code according \cite{PhysRevD.96.044028}
\begin{align}
\mathcal{F}=\frac{1}{16\pi}\sum_\ell\sum_{m=-\ell}^{\ell}|\dot{h}_{\ell m}|^2,
\end{align}
where $h_{\ell m}$ are spherical harmonic modes of the gravitational waves.

Regarding to numerical implementation, we need note one fact. Near merger, $\dot{p}_{r}$ becomes very large which strongly affects the numerical accuracy. To treat this problem we follow the trick used by usual EOB code \cite{PhysRevD.81.084041} to replace the original canonical momentum $p_r$ with the momentum of the tortoise coordinates $p_{r*}$. The detail replacement relations are
\begin{align}
    &p_r=p_{r*}\left(1+\frac{2}{r^2c^2}+\frac{4-3\nu}{r^2c^4}\right), \\
    &\dot{p}_{r}=\dot{p}_{r*}-\frac{2(p_{r*}^2-r\dot{p}_{r*})}{r^2c^2}+\frac{1}{r^3c^4}\left[r(r\dot{p}_{r*}(4-3\nu) \right.\nonumber \\
    &\left.+rp^4_{r*}(1+\nu)+p^2_{r*}(5(\nu-1)+r^2\dot{p}_{r*}(1+\nu))\right].
\end{align}

Besides the dynamical equations we have described above, another important issue about dynamics is the initial conditions. Adiabatically the spin-precession orbits can be looked as evolved Keplerian orbits. So we need only to determine the initial Keplerian orbit to quantify the initial condition. The initial Keplerian orbit can be described by mean orbital angular velocity $f_0$ and eccentricity $e_0$. Similar to what we have done in \cite{PhysRevD.96.044028}, we solve
\begin{align}
\pdv{H_{\text{EOB}}}{p_\phi}&=\frac{\pi}{f_0}, \\
\pdv{H_{\text{EOB}}}{r}&=-\frac{e_0}{r^2},
\end{align}
for the initial conditions. Here we call $f_0$ the initial orbital angular velocity and $e_0$ the initial eccentricity.
The remaining steps follow those proposed in \cite{PhysRevD.74.104005}. There are several different definitions of eccentricity for binary system in general relativity \cite{2019CQGra..36b5004L,2022ApJ...936..172K}. Many definitions are gauge dependent. In the current work, we do not pay attention to the concept subtlety of eccentricity. In stead we just take it as a parameter to describe the waveform.
\section{Factorized waveform}\label{sec3}
Following \cite{PhysRevD.89.084006} we calculate gravitational waveform for spin-precessing BBHs in the precession frame. Similar to \cite{seobnrv5}, we define three mutually orthogonal normal vectors $(\pmb{n},\pmb{\lambda},\pmb{e})$ via
\begin{align}
&\pmb{n}:=\frac{\pmb{r}}{r}, \\
&\pmb{e}:=\frac{\pmb{L}_N}{L_N}, \\
&\pmb{\lambda}:=\pmb{e}\times\pmb{n},
\end{align}
where $\pmb{L}_N:=\pmb{r}\times\pmb{v}$, with $\pmb{v}:=\dot{\pmb{r}}$ is velocity. These vectors constitute a coordinate system in the precession frame.

Up to 2PN order, the gravitational wave waveform for eccentric spin-precessing BBHs in harmonic coordinates has been calculated in \cite{PhysRevD.54.4813} for orbital motion part and in \cite{PhysRevD.87.044009} for spin-orbit part. We then transform the harmonic coordinate to EOB coordinate according \cite{PhysRevD.104.024046}. Then we apply a rotation to the waveform to get the waveform in comoving frame. The rotation is the one to let $\pmb{r}$ and $\pmb{p}$ lie in x-y plane. After that we decompose the waveform in comoving frame into spin-weighted -2 spherical harmonic modes. Finally we express the results in terms of $r$, $p_r$, $L$ and
\begin{align}
\chi_{(n,\lambda,e)}\equiv\pmb{\chi}\cdot (\pmb{n},\pmb{\lambda},\pmb{e}).
\end{align}
The detail expressions are presented in the supplementary material. Our waveform can recover the results of \cite{PhysRevD.104.024046} in the spin aligned binary cases.

Next we re-sum the waveform into the factorized form \cite{PhysRevD.83.064003,PhysRevD.79.064004,PhysRevD.102.044055,PhysRevD.104.024046}
\begin{align}
h^F_{\ell m}=h^N_{\ell m}\hat{S}_{\text{eff}}T^{\text{qc}}_{\ell m}e^{i\delta_{\ell m}}f_{\ell m},\label{eq3}
\end{align}
where $h^N_{\ell m}$ is the Newtonian order of the waveform. We adopt the form used in \cite{PhysRevD.102.044055} for $T^{\text{qc}}_{\ell m}e^{i\delta_{\ell m}}$ to count the tail contributions.
The effective source term $\hat{S}_{\text{eff}}$ is given by
\begin{align}
\hat{S}_{\text{eff}}=
\begin{cases}
\hat{H}_{\text{eff}} & \text{$\ell+m$ is even} \\
v_\Omega L & \text{$\ell+m$ is odd}
\end{cases}.
\end{align}
In \cite{PhysRevD.81.084024,PhysRevD.84.104027}, $\hat{H}_{\text{eff}}$ has been given in terms of $r$, $p_r$, $L$, $\chi_{1,2n}$, $\chi_{1,2\lambda}$ and $\chi_{1,2e}$. In the current work we adopt this function form but replace $L$ with $\beta\equiv\sqrt{1+r^2\dot{p}_r}$. In addition we expand $v_\Omega\equiv\Omega^{1/3}$ and $L$ in terms of $r$, $p_r$, and $\beta$ through relations (Eq.~(20) of \cite{PhysRevD.104.024046})
\begin{widetext}
\begin{align}
&\frac{L^2}{r}=\beta^2+\frac{1}{2rc^2}\left[\beta ^4 (\nu +1)-3 \beta ^2 (\nu -1)+2 (\nu +1)+r \left(\beta ^2 (\nu +1)-\nu +3\right) p_r^2\right] +\frac{3 \beta  \left(\chi_{se} (\nu -2)-2 \chi_{ae} \sqrt{1-4 \nu }\right)}{c^3r^{3/2}} \\
&+\frac{1}{8 r^2c^4}\left[\beta ^6 \nu ^2+5 \beta ^6 \nu +\beta ^6-3 \beta ^4 \nu ^2-5 \beta ^4 \nu +13 \beta ^4+2 \beta ^2 \nu ^2-12 \beta ^2 \nu +34 \beta ^2+48 \delta  \chi_{a\lambda} \chi_{s\lambda} \nu -64 \beta ^2 \chi_{ae}^2 \nu  \right.\nonumber \\
&\left.+16 \beta ^2 \chi_{ae}^2+2 r p_r^2 \left(3 \beta ^2+\nu  \left(3 \beta ^4+\beta ^2 (\nu -5)+16 \chi_{a\lambda}^2-16 \chi_{s\lambda}^2 (\nu -1)-\nu \right)+4 \chi_{ae}^2 (4 \nu -1)-4 \chi_{se}^2 (1-2 \nu )^2 \right.\right.\nonumber \\
&\left.\left.-4 \chi_{a\lambda}^2-4 \chi_{s\lambda}^2-22 \nu +5\right)-64 \beta ^2 \delta  \chi_{ae} \chi_{se} \nu +32 \beta ^2 \delta  \chi_{ae} \chi_{se}+48 \delta  \chi_{ae} \chi_{se} \nu +32 \delta  \chi_{ae} \chi_{se} \nu  r p_r^2-16 \delta  \chi_{ae} \chi_{se} r p_r^2\right.\nonumber\\
&\left.+64 \beta ^2 \chi_{se}^2 \nu ^2-64 \beta ^2 \chi_{se}^2 \nu +16 \beta ^2 \chi_{se}^2-48 \chi_{se}^2 \nu ^2+48 \chi_{se}^2 \nu -48 \chi_{s\lambda}^2 \nu ^2+48 \chi_{s\lambda}^2 \nu -12 \nu -8 \left(2 \beta ^2-3\right) (4 \nu -1) \chi_{an}^2 \right.\nonumber\\
&\left.-16 \delta  \chi_{an} \chi_{sn} \left(\beta ^2 (4 \nu -2)+3\right)+8 \chi_{sn}^2 \left(2 \beta ^2 (1-2 \nu )^2-3\right)-\left(\beta ^2-1\right) ((\nu -1) \nu +1) r^2 p_r^4+32 \delta  \chi_{a\lambda} \chi_{s\lambda} \nu  r p_r^2 \right.\nonumber\\
&\left.-16 \delta  \chi_{a\lambda} \chi_{s\lambda} r p_r^2+24\right], \\
%
% v_\Omega = \Omega^1/3
%
&\sqrt{r}v_\Omega=\beta^{1/3}+\frac{1}{12r\beta^{5/3}c^2}\left[-\left(\beta ^4 (\nu +1)\right)+\beta ^2 (\nu -1)+2 (\nu +1)-r \left(\beta ^2 (\nu +1)+\nu -3\right) p_r^2\right]+\frac{\chi_{se} (\nu -2)-2 \delta  \chi_{ae}}{6 \beta ^{2/3} r^{3/2}c^3} \\
&+\frac{1}{288 \beta ^{11/3} r^2}\left[\beta ^8 ((\nu -16) \nu +1)-2 \beta ^6 (\nu  (4 \nu -9)+14)-7 \beta ^4 (\nu +3) (3 \nu +1)+2 r p_r^2 \left(\beta ^6 (\nu  (7 \nu -4)+7)\right.\right.\nonumber\\
&\left.\left.+\beta ^4 \left(-6 \nu ^2+4 \nu -8\right)+\beta ^2 \left(96 \delta  \chi_{a\lambda} \chi_{s\lambda} \nu -48 \delta  \chi_{a\lambda} \chi_{s\lambda}+24 \chi_{ae}^2 (4 \nu -1)+48 \delta  \chi_{ae} \chi_{se} (2 \nu -1)-24 \chi_{se}^2 (1-2 \nu )^2\right.\right.\right.\nonumber\\
&\left.\left.\left.+24 \chi_{a\lambda}^2 (4 \nu -1)-96 \chi_{s\lambda}^2 \nu ^2+96 \chi_{s\lambda}^2 \nu -24 \chi_{s\lambda}^2-39 \nu ^2-84 \nu -41\right)+10 (\nu -3) (\nu +1)\right)\right.\nonumber\\
&\left.+4 \beta ^2 \left(\nu  \left(72 \left(\delta  \chi_{a\lambda} \chi_{s\lambda}+\delta  \chi_{ae} \chi_{se}-\nu  \left(\chi_{se}^2+\chi_{s\lambda}^2\right)+\chi_{se}^2+\chi_{s\lambda}^2\right)+19 \nu -18\right)+36 (4 \nu -1) \chi_{an}^2-72 \delta  \chi_{an} \chi_{sn}\right.\right.\nonumber\\
&\left.\left.-36 \chi_{sn}^2+17\right)-20 (\nu +1)^2+r^2 \left(\beta ^4 (\nu  (13 \nu +8)+13)+\beta ^2 \left(20 \nu ^2-34 \nu -36\right)-5 (\nu -3)^2\right) p_r^4\right].
\end{align}

So based on the re-summation relation we can determine $f_{\ell m}$. And more we divide the got $f_{\ell m}$ into two parts. One is the quasi-circular part corresponding to $p_r=\dot{p}_r=0$. The another is the elliptical correction part $f^e_{\ell m}$ corresponding to the left part other than $p_r=\dot{p}_r=0$. For odd $m$ cases, the quasi-circular part is divided into spinless part $\rho^\ell_{\ell m}$ and spin part $f^s_{\ell m}$ more \cite{PhysRevD.104.024046}. Consequently we have
\begin{align}
&\qquad f_{\ell m}=\begin{cases}
\rho^\ell_{\ell m}(r,\beta,\chi_{1,2n},\chi_{1,2\lambda},\chi_{1,2e})+f^e_{\ell m}(r,p_r,\beta,\chi_{1,2n},\chi_{1,2\lambda},\chi_{1,2e}) & \text{$m$ is even} \\
\rho^\ell_{\ell m}(r,\beta)+f^s_{\ell m}(r,\beta,\chi_{1,2n},\chi_{1,2\lambda},\chi_{1,2e})+f^e_{\ell m}(r,p_r,\beta,\chi_{1,2n},\chi_{1,2\lambda},\chi_{1,2e}) &\text{$m$ is odd}
\end{cases}.
\end{align}

We rearrange the auto-variables for the quasi-circular part to form $v_{\Omega}$. So we can express $f_{\ell m}$ as
\begin{align}
&\qquad f_{\ell m}=\begin{cases}
\rho^\ell_{\ell m}(v_\Omega,\chi_{1,2n},\chi_{1,2\lambda},\chi_{1,2e})+f^e_{\ell m}(r,p_r,\beta,\chi_{1,2n},\chi_{1,2\lambda},\chi_{1,2e}) & \text{$m$ is even} \\
\rho^\ell_{\ell m}(v_\Omega)+f^e_{\ell m}(r,p_r,\beta,\chi_{1,2n},\chi_{1,2\lambda},\chi_{1,2e})+f^s_{\ell m}(v_\Omega,\chi_{1,2n},\chi_{1,2\lambda},\chi_{1,2e}) &\text{$m$ is odd}
\end{cases}.
\end{align}
\end{widetext}

\begin{figure*}[t]
\centering
\begin{tabular}{c}
\includegraphics[width=\textwidth]{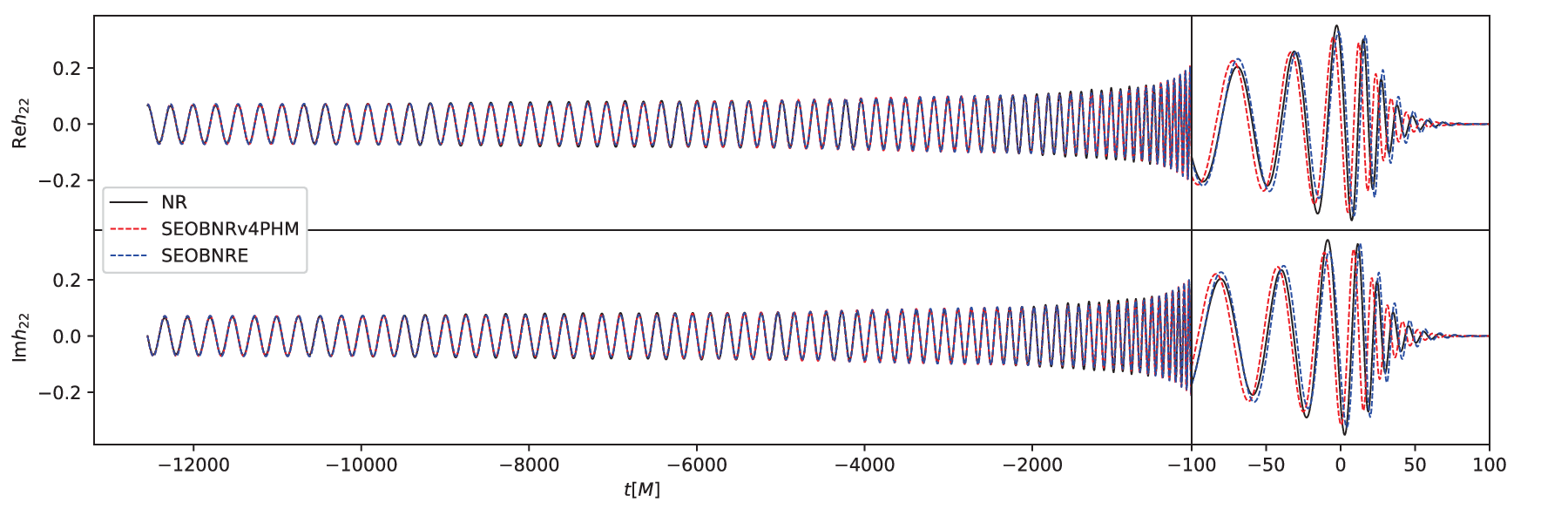}
\end{tabular}
\caption{Waveform comparison for $(2,2)$ mode between numerical relativity and EOB models including newly got $\texttt{SEOBNRE}$ model in the current paper and $\texttt{SEOBNRv4PHM}$. The top row is the real part of the $(2,2)$ mode. The bottom row is the imaginary part of the $(2,2)$ mode. SXS:BBH:1106 (mass ratio $q=1.68$, BH's spin $\vec{\chi}_1=(0.72,-0.00,0.36)$ and $\vec{\chi}_2=(-0.14,-0.07,0.05)$) \cite{SXSBBH} case is used in this plot.}
\label{fig1}
\end{figure*}

\begin{figure*}[t]
\centering
\begin{tabular}{c}
\includegraphics[width=\textwidth]{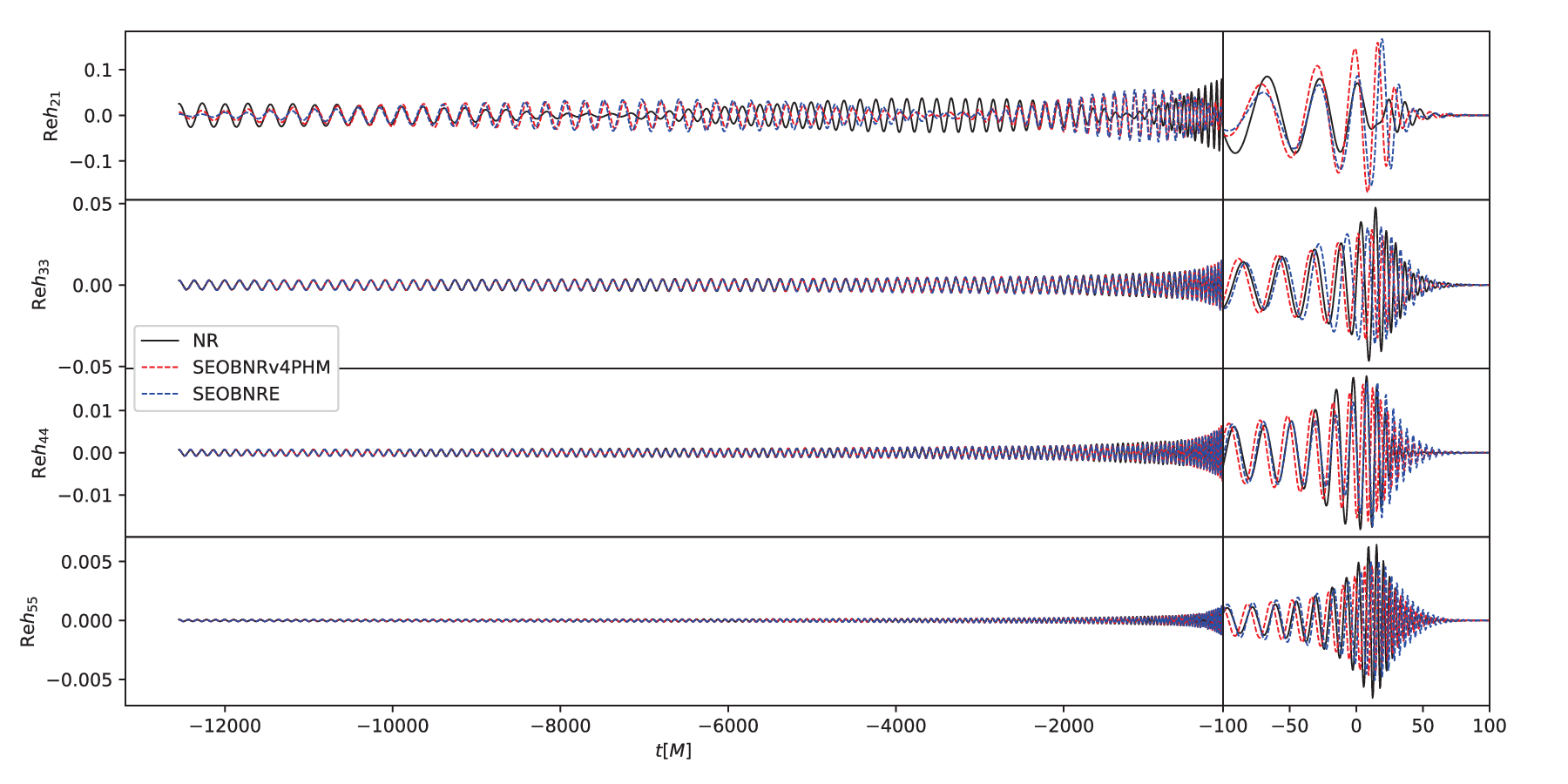}
\end{tabular}
\caption{Waveform comparison for the real parts of higher modes $(2,1)$, $(3,3)$, $(4,4)$ and $(5,5)$ between numerical relativity and EOB models including newly got $\texttt{SEOBNRE}$ model in the current paper and $\texttt{SEOBNRv4PHM}$. Similar to Fig.~\ref{fig1}, SXS:BBH:1106 (mass ratio $q=1.68$, BH's spin $\vec{\chi}_1=(0.72,-0.00,0.36)$ and $\vec{\chi}_2=(-0.14,-0.07,0.05)$) \cite{SXSBBH} case is used in this plot.}
\label{fig2}
\end{figure*}

Comparing to the waveform used in \cite{PhysRevD.104.024046}, we introduce the extra contribution of the BH's spin component perpendicular to the orbital angular momentum in the circular orbital part $\rho_{\ell m}$ for even-parity modes and spin contribution $f^s_{\ell m}$ for odd-parity modes. We list the detail expressions for these extra terms in Appendix~\ref{appB}. The elliptical corrections $f^e_{\ell m}$ are more complicated. We present them in the supplementary material.

Note that the waveform (\ref{eq3}) is respect to the precession frame. In order to get the waveform in inertial frame we firstly calculate the state variables according to the dynamical equations described in the last section. From the calculated state variables we can get $r$, $\beta$, $\chi_{1,2n}$, $\chi_{1,2\lambda}$ and $\chi_{1,2e}$. Consequently we can calculate the waveform in the precession frame using (\ref{eq3}). After that we apply an inverse rotation operation on the waveforms $h_{\ell m}$ \cite{PhysRevD.89.084006}. The rotation operation is the one lets $\pmb{r}$ and $\pmb{p}$ lie in x-y plane.
\section{Performance of the spin-precession $\texttt{SEOBNRE}$ waveform model}\label{sec4}

Along our series work about $\texttt{SEOBNRE}$ waveform models \cite{PhysRevD.96.044028,PhysRevD.101.044049_validSEOBNRE,Liu_2022,2023IJMPD..3250015L}, the current work is the first one describing spin-precession. Before considering the spin-precession effect, we have checked the performance of our new $\texttt{SEOBNRE}$ waveform model for the spin-aligned cases. We find our new model can achieve as good performance as the last version $\texttt{SEOBNRE}$ waveform model \cite{Liu_2022}. Compared to the last version $\texttt{SEOBNRE}$ waveform model \cite{Liu_2022}, the (5,5) mode is newly constructed. In Appendix~\ref{appnew}, we show the test of the spin-precession $\texttt{SEOBNRE}$ waveform model against the spin-aligned BBHs. Regarding to spin-precession, we would like to firstly check the performance of our waveform model for circular spin-precessing cases. After that we check generic eccentric spin-precessing cases. Then we quantify the performance of our waveform model with matching factor which has been widely used to estimate the accuracy of a given waveform model. Specifically SXS numerical relativity waveform catalog and RIT numerical relativity waveform catalog are used to do the waveform comparison and matching factor calculation.
\subsection{Circular spin-precessing cases}

\begin{figure*}[t]
\centering
\begin{tabular}{c}
\includegraphics[width=\textwidth]{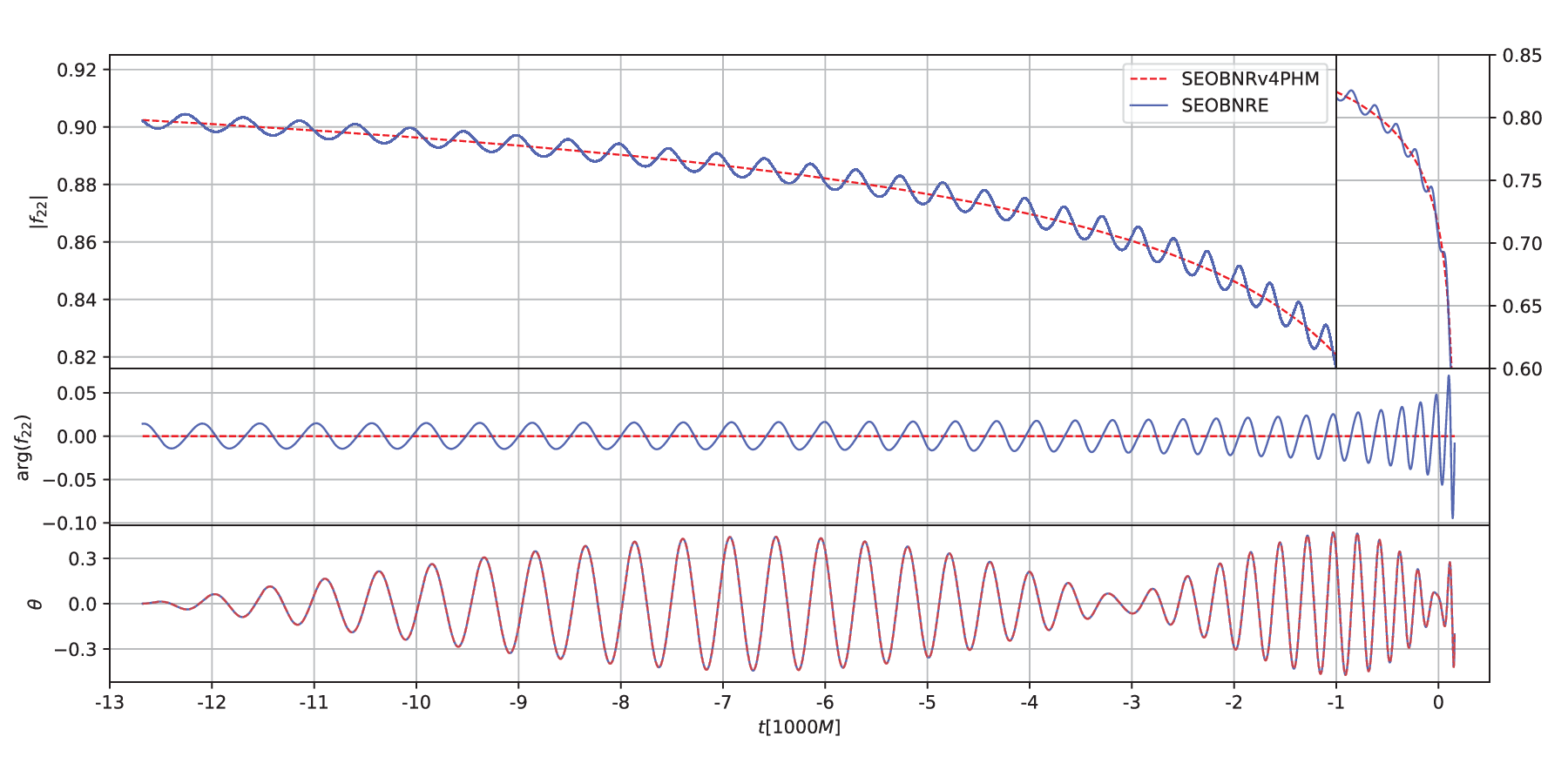}
\end{tabular}
\caption{Corresponding to Fig.~\ref{fig1} and \ref{fig2}, the orbital plan precession behavior (bottom panel) and the waveform correction factor $f_{22}$ (top panel: amplitude of $f_{22}$, middle panel: phase of $f_{22}$) are plotted. Comparison between $\texttt{SEOBNRv4PHM}$ waveform model and $\texttt{SEOBNRE}$ waveform model is shown.}
\label{fig3}
\end{figure*}

$\texttt{SEOBNRv4PHM}$ waveform model is valid for circular spin-precession cases. In order to check the performance of our new waveform model, we can compare our resulted waveform to both the $\texttt{SEOBNRv4PHM}$ waveform and the numerical relativity waveform. As an example, we randomly choose a circular spin-precession waveform from SXS catalog \cite{SXSBBH}, SXS:BBH:1106 which has mass ratio $q=1.68$ and spin $\vec{\chi}_1=(0.72,-0.00,0.36)$ $\vec{\chi}_2=(-0.14,-0.07,0.05)$. For the EOB waveforms we set the parameters of mass ratio and BH's spin accordingly. For comparison, we align the time of the waveforms with setting the corresponding time of maximal $|h_{22}|$ as 0. After that we adjust the initial phase of EOB waveforms to match the NR waveform. Then we plot the results in Fig.~\ref{fig1}. From this comparison we can see that both $\texttt{SEOBNRv4PHM}$ waveform model and our newly developed $\texttt{SEOBNRE}$ waveform model can recover the NR results quite well. And more we can see that the real part and imaginary part show similar behavior. Such similar behavior happens for all cases we have checked and for all waveform modes. So in the following we just plot real part. In Fig.~\ref{fig2} we compare the higher modes for $(2,1)$, $(3,3)$, $(4,4)$ and $(5,5)$. The consistency between NR waveforms and EOB models can be seen there.

\begin{figure*}[t]
\centering
\begin{tabular}{c}
\includegraphics[width=\textwidth]{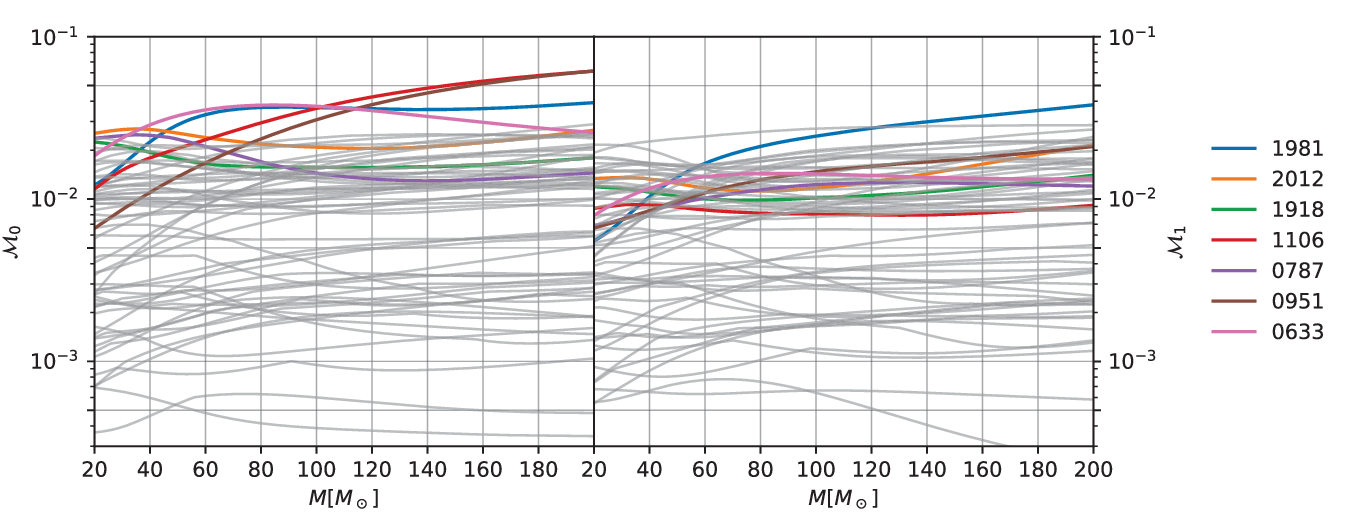}
\end{tabular}
\caption{Mismatch factor of $(2,2)$ spherical mode between NR waveform and EOB model waveforms. The left panel is for $\texttt{SEOBNRv4PHM}$ model. The right panel is for $\texttt{SEOBNRE}$ model. The Advanced LIGO designed sensitivity is used in this figure.}
\label{fig4}
\end{figure*}

Phenomenologically spin-precession makes the binary orbit plane precess. If we initially set the orbit plane coincide with x-y plane, such precession behavior shows variation of EOB spherical coordinate $\theta$. Using SXS:BBH:1106 as an example, we investigate such plane precession behavior respect to time. The result is plotted in bottom panel of Fig.~\ref{fig3}. The oscillation behavior of $\theta$ is clear. Along with this $\theta$ we also plot the model waveform correction factor $f_{22}$ described in previous section. For comparison, both factors for $\texttt{SEOBNRv4PHM}$ waveform model and $\texttt{SEOBNRE}$ waveform model. Interestingly, we find that the factor of $\texttt{SEOBNRE}$ waveform model can closely catch up the oscillation behavior of $\theta$. In contrast, the factor of $\texttt{SEOBNRv4PHM}$ waveform model does not count such oscillation at all. We understand this fact as following. $\texttt{SEOBNRv4PHM}$ waveform model neglects the contribution of the non-perpendicular spin by assuming that such contribution is small. Differently $\texttt{SEOBNRE}$ waveform model faithfully counts the contribution of the non-perpendicular spin. It is this non-perpendicular spin that introduces the oscillation behavior of $f_{22}$. In fact other correction factors including $f_{\ell m}$ all admit this oscillation behavior in the $\texttt{SEOBNRE}$ waveform model.

In order to quantitatively check the consistency between the NR waveform and EOB model waveforms, we calculate the mismatch factors. Due to initial oscillations in NR simulations, when computing the best-matching EOB waveform, we search for a waveform that matches well in the vicinity of the initial frequency given by NR, serving as the best-matching waveform. We define the mismatch factor $\mathcal{M}$ between two waveforms, $h_1$ and $h_2$ via
\begin{align}
\mathcal{M} = 1 - \max_{t_c,\phi_c}{\frac{\ip{h_1}{h_2}}{\sqrt{\ip{h_1}\ip{h_2}}}},
\end{align}
where the inner product $\ip{\cdot}{\cdot}$ is given by
\begin{align}
\ip{h_1}{h_2}=\int_{f_{\text{min}}}^{f_{\text{max}}}\frac{\tilde{h}_1(f)\tilde{h}_2^*(f)}{S_n(f)}\dd{f}.
\end{align}
We use the power spectrum density (PSD) of the high power designed Advanced LIGO \cite{advLIGOPSD} as the one sided power spectrum of detectors noise $S_n$. We compute the mismatch factor $\mathcal{M}$ between the (2,2) mode of the numerical relativity waveforms and the EOB waveforms. This mismatch factor $\mathcal{M}$ depends on the total mass $M$ of the binary black hole system. The chosen range for the total mass is from 20 to 200 solar masses, which roughly corresponds to the detection range of ground-based detectors. The best-matching waveform is the one that achieves the highest overall matching in the selected mass range.

In Fig.~\ref{fig4}, we use $\mathcal{M}_0$ to denote the mismatch factor between the NR waveform and the $\texttt{SEOBNRv4PHM}$ model waveform and use $\mathcal{M}_1$ to denote the mismatch factor between the NR waveform and the $\texttt{SEOBNRE}$ model waveform. The plot shows the mismatch factor as a function of the BBH total mass. In all 63 NR waveforms are used here. We can see that $\texttt{SEOBNRE}$ model behaves as good as $\texttt{SEOBNRv4PHM}$ model in most cases. In some cases, $\texttt{SEOBNRE}$ model can even show a little bit better behavior. This small improvement, about $10^{-2}$, is due to the non-perpendicular spin contribution afore mentioned.

\subsection{Eccentric spin-precessing cases}
In the previous subsection, we have investigated the behavior of $\texttt{SEOBNRE}$ waveform model in the circular orbit spin-precessing cases. In this subsection we move on to check the behavior of $\texttt{SEOBNRE}$ waveform model for general cases with both spin precession and orbit eccentricity. To the best of our knowledge, our $\texttt{SEOBNRE}$ waveform model is the first theoretical waveform model to treat such generic BBHs.

\begin{figure*}[t]
\centering
\begin{tabular}{c}
\includegraphics[width=\textwidth]{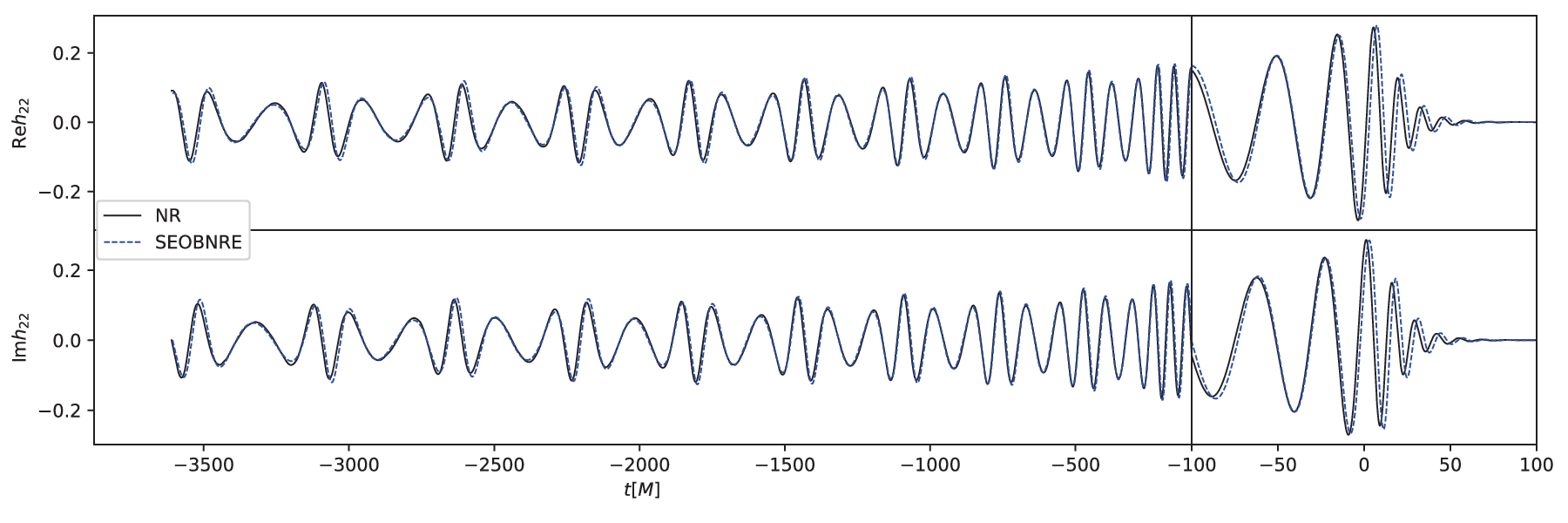}
\end{tabular}
\caption{Waveform comparison for $(2,2)$ mode between numerical relativity and $\texttt{SEOBNRE}$ model. The top row is the real part of the $(2,2)$ mode. The bottom row is the imaginary part of the $(2,2)$ mode. RIT:eBBH:1632 (mass ratio $q=1$, BH's spin $\vec{\chi}_1=(0.7,0,0)$ and $\vec{\chi}_2=(0.7,0,0)$) \cite{PhysRevD.105.124010} case is used in this plot.}
\label{fig5}
\end{figure*}

\begin{figure*}[t]
\centering
\begin{tabular}{c}
\includegraphics[width=\textwidth]{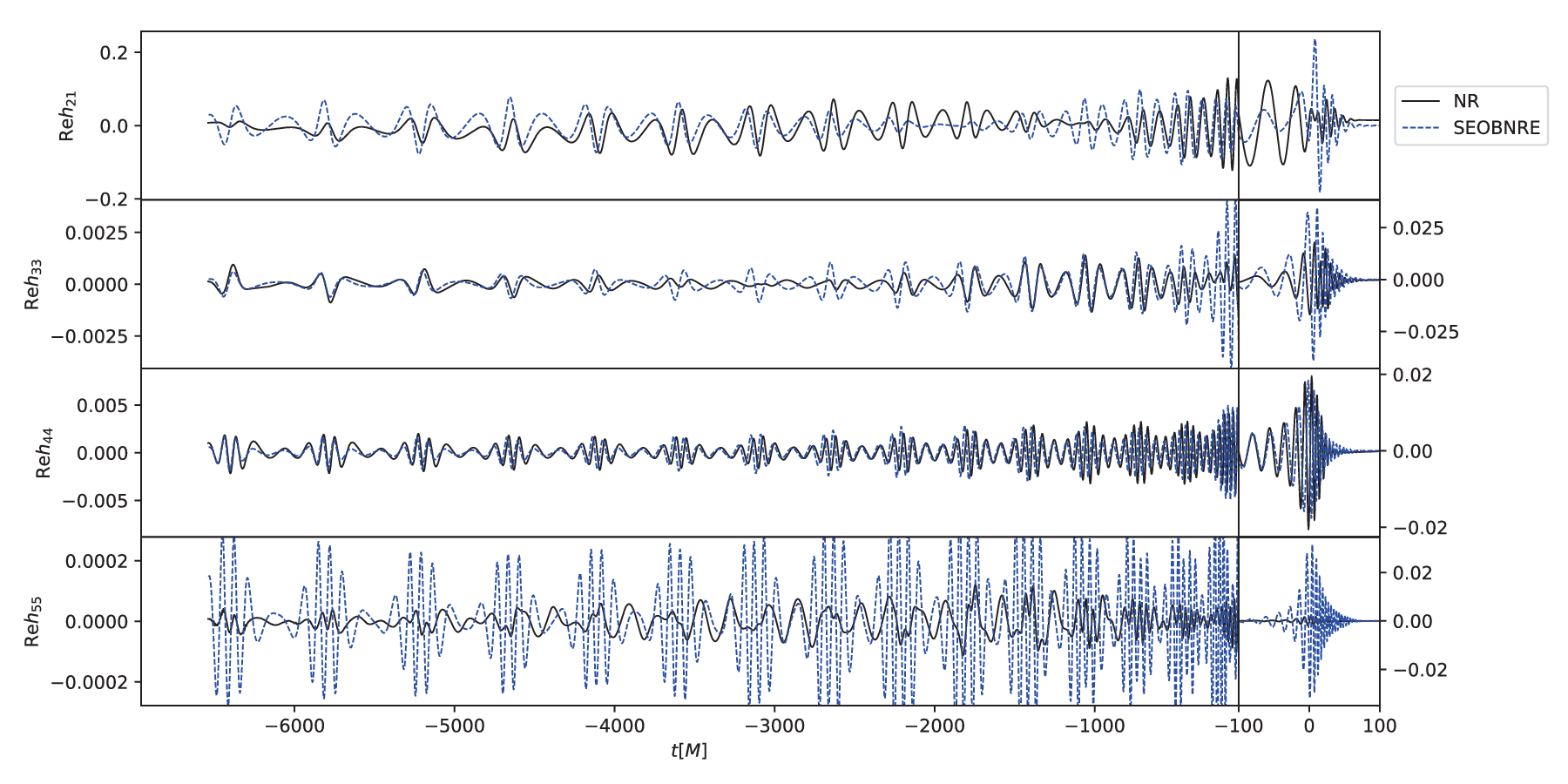}
\end{tabular}
\caption{Waveform comparison for the real parts of higher modes $(2,1)$, $(3,3)$, $(4,4)$ and $(5,5)$ between numerical relativity and $\texttt{SEOBNRE}$ model. Similar to Fig.~\ref{fig5}, RIT:eBBH:1632 (mass ratio $q=1$, BH's spin $\vec{\chi}_1=(0.7,0,0)$ and $\vec{\chi}_2=(0.7,0,0)$) \cite{PhysRevD.105.124010} case is used in this plot.}
\label{fig6}
\end{figure*}

Since SXS catalog has not simulation results for general BBH with both spin precession and orbit eccentricity, we adopt RIT catalog instead \cite{PhysRevD.105.124010}. As an example, we chose RIT:eBBH:1632 to compare the waveform between the NR and the $\texttt{SEOBNRE}$ model. The mass ratio is 1, the initial spin configuration is $\vec{\chi}_1=(0.7,0,0), \vec{\chi}_2=(0.7,0,0)$. The initial eccentricity is $0.2775$ at $Mf_{22}=0.002463$. Numerical relativistic simulations exhibit oscillations in the initial stage. After drop initial part NR waveform, we simultaneously search for the initial frequency $f_0$ and initial eccentricity $e_0$ through comparing the NR waveform to the $\texttt{SEOBNRE}$ model waveforms.

\begin{figure*}[t]
\centering
\begin{tabular}{c}
\includegraphics[width=\textwidth]{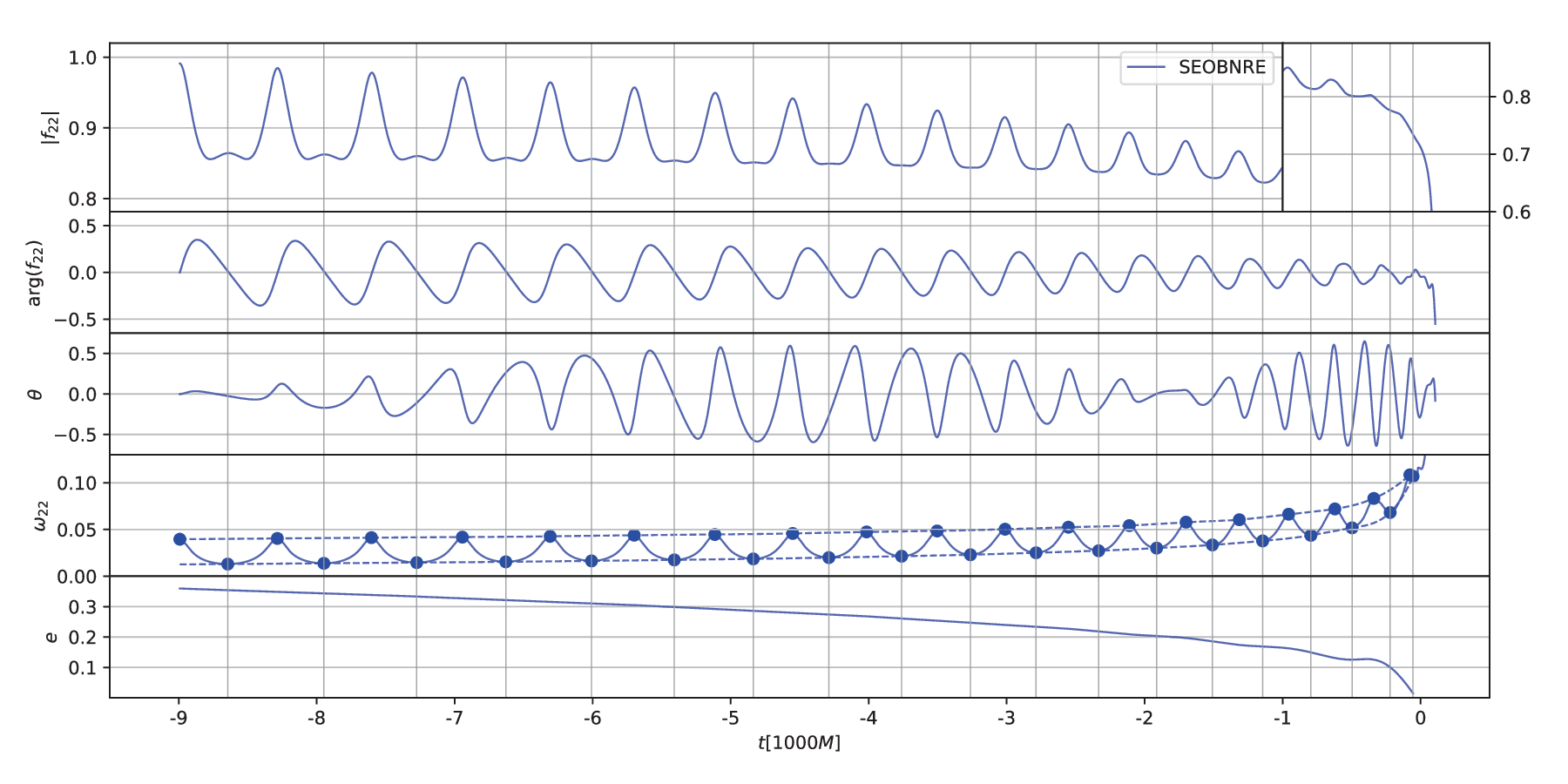}
\end{tabular}
\caption{Similar to Fig.~\ref{fig3} and corresponding to Figs.~\ref{fig5} and \ref{fig6}, the $\texttt{SEOBNRE}$ model waveform correction factor $f_{22}$ (top panel: amplitude of $f_{22}$, middle panel: phase of $f_{22}$) are plotted. Along with that, the orbit plane precession behavior of $\theta$ is plotted in the third panel. The fourth panel is the model waveform frequency of (2,2) mode $\omega_{22}$ corresponding to the co-precession frame. The fifth panel if the calculated eccentricity according to $\omega_{22}$ shown in the fourth panel.}
\label{fig7}
\end{figure*}
We extend the method proposed in \cite{2023arXiv230211257A} to estimate the eccentricity $e_{\rm gw}(t)$ based on the gravitational waveform. Since there are spin-precession effects in our case, we calculate the eccentricity in the co-precession frame. The eccentricity of gravitational waves is defined by
\begin{align}
e_{\rm gw}=\cos{(\Psi/3)} - \sqrt{3}\sin{(\Psi/3)},
\end{align}
where
\begin{align}
\Psi:=\arctan{\left(\frac{1-e^2_{\omega_{22}}}{2e_{\omega_{22}}}\right)}.
\end{align}
The $e_{\omega_{22}}$ represent the generalized eccentricity that defined by the frequency of the dominant $(2,2)$ mode $\omega_{22}$,
\begin{align}
e_{\omega_{22}}&=\frac{\sqrt{\omega^p_{22}}-\sqrt{\omega^a_{22}}}{\sqrt{\omega^p_{22}}+\sqrt{\omega^a_{22}}},
\end{align}
where the superscripts $p,a$, denotes the frequency that evaluated at pericenters and apocenters.

In Fig.~\ref{fig5}, we present the comparison results between the best-matching $\texttt{SEOBNRE}$ model waveform and the NR waveform. The consistency between these two waveforms is clear. This figure is similar to Fig.~\ref{fig1} but for general cases with both spin precession and orbit eccentricity. Similar to Fig.~\ref{fig2} we plot the higher modes in Fig.~\ref{fig6}. Similar to Fig.~\ref{fig3}, we check the model waveform correction factor $f_{22}$ and correspondingly the orbit plane precession and the evolution behavior of eccentricity in Fig.~\ref{fig7}. Besides the behavior of $f_{22}$, we have also plotted the orbit plane precession behavior of $\theta$, the model waveform frequency of (2,2) mode $\omega_{22}$ corresponding to the co-precession frame and the calculated eccentricity according to the got $\omega_{22}$. In addition to the oscillation behavior of $f_{22}$ resulted from the orbit plan precession, we can also see another oscillation behavior which happens when the orbit passes by the pericenters. Interestingly there is a local maximal of $|f_{22}|$ happening exactly when the orbit locates at the pericenters. Not like the spin-aligned cases, the eccentricity may be effected by the orbit precession. During the evolution process, when the orbit precession becomes strongly enough, the orbit eccentricity shows oscillation behavior as shown in the bottom panel of Fig.~\ref{fig7}, near merger.
\begin{figure}
\centering
\begin{tabular}{c}
\includegraphics[width=0.5\textwidth]{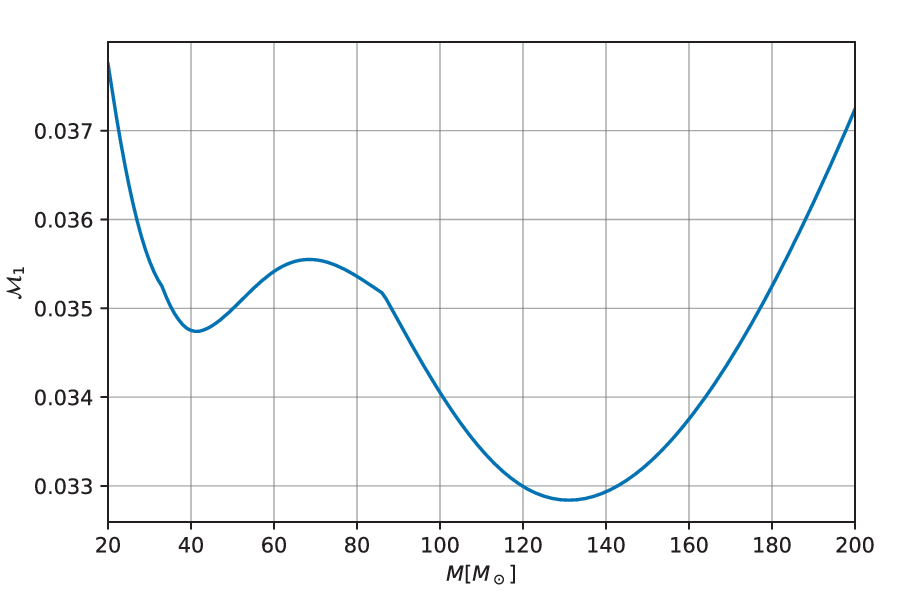}
\end{tabular}
\caption{Mismatch factor of $(2,2)$ spherical mode between the NR waveform and the $\texttt{SEOBNRE}$ model waveform. Similar to Fig.~\ref{fig5}, RIT:eBBH:1632 (mass ratio $q=1$, BH's spin $\vec{\chi}_1=(0.7,0,0)$ and $\vec{\chi}_2=(0.7,0,0)$) \cite{PhysRevD.105.124010} case is investigated in this plot.}
\label{fig8}
\end{figure}

In order to quantitatively estimate the accuracy of the $\texttt{SEOBNRE}$ model waveform, we again calculate the mismatch factor between the model waveform and the NR waveform. We plot the result in Fig.~\ref{fig8}. Using RIT:eBBH:1632 (mass ratio $q=1$, BH's spin $\vec{\chi}_1=(0.7,0,0)$ and $\vec{\chi}_2=(0.7,0,0)$) \cite{PhysRevD.105.124010} as an example of generic eccentric spin-precession BBH, we find that the matching factor achieved within the mass range of 20 to 200 solar masses better than 96.2\%.

Comparing Fig.~\ref{fig6} to Fig.~\ref{fig2} and \ref{fig12}, we can find that the higher modes for eccentric spin-precession case perform worse than individual eccentric case and spin-precession circular case. This means our model has not treated the interaction between eccentricity and spin-precession well. But the good news is the matching factor of dominating (2,2) mode between NR waveform and $\texttt{SEOBNRE}$ model waveform is better than 96\% as we have seen in Fig.~\ref{fig8}.

\begin{figure*}[t]
\centering
\begin{tabular}{c}
\includegraphics[width=\textwidth]{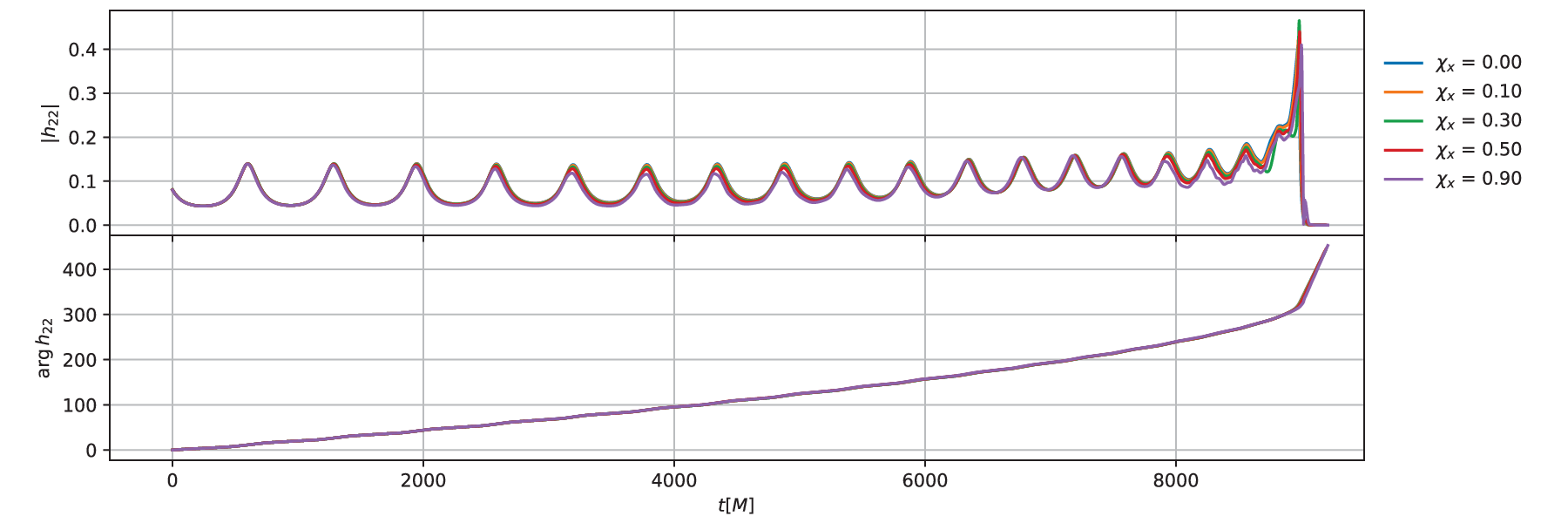}
\end{tabular}
\caption{Waveforms comparison for different spin amplitudes. Corresponding to RIT:eBBH:1632, mass ratio $q=1$ and initial eccentricity $e_0=0.37$ at $Mf_{22}=0.003$ are used in this figure. Different lines correspond to different BHs' spin amplitude $\chi_x$. The BHs' spins are set as $\vec{\chi}_1=(\chi_x,0,0)$ and $\vec{\chi}_2=(\chi_x,0,0)$.}
\label{fig9}
\end{figure*}
\begin{figure*}
\centering
\begin{tabular}{c}
\includegraphics[width=\textwidth]{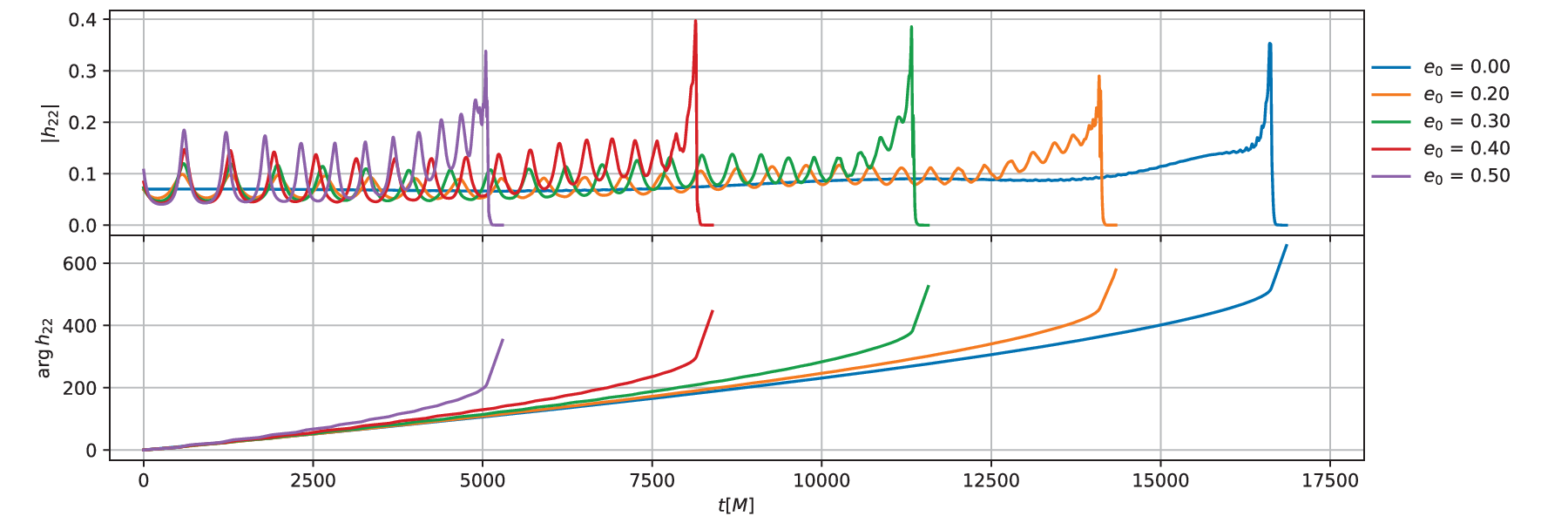}
\end{tabular}
\caption{Waveforms comparison for different initial orbit eccentricity $e_0$. Corresponding to RIT:eBBH:1632, mass ratio $q=1$ and BH's spin $\vec{\chi}_1=(0.7,0,0)$ and $\vec{\chi}_2=(0.7,0,0)$ are used in this figure. Different lines correspond to different initial orbit eccentricity at $Mf_{22}=0.003$.}
\label{fig10}
\end{figure*}
Now we can use $\texttt{SEOBNRE}$ waveform model as tool to investigate the effect of spin precession and orbit eccentricity individually on the waveform. We use RIT:eBBH:1632 as a reference but change BHs' spin or change orbit eccentricity to check the waveform behavior.

In Fig.~\ref{fig9} we investigate the effect of BH spin amplitude on the waveform. The effect of spin on waveform comes in at relative high post-Newtonian order. Consequently the waveforms corresponding to different spin amplitude are similar to each other. In the current spin-precession cases, larger spin results in larger precession. We can consequently see precession correction to the waveform amplitude at about time $t=4000M$ and $t=8000M$. Interestingly we find that the correction to the waveform phase is small.

In Fig.~\ref{fig10} we investigate the effect of orbit eccentricity on the waveform. As ones expected, larger eccentricity results in faster inspiral and merger. In addition, larger eccentricity results in stronger waveform burst behavior. Consequently the waveform phase for larger eccentricity shows faster increasing behavior. From Fig.~\ref{fig10} we can see different initial eccentricity gives clearly different behavior.
\begin{figure*}
\centering
\begin{tabular}{cc}
\includegraphics[width=0.5\textwidth]{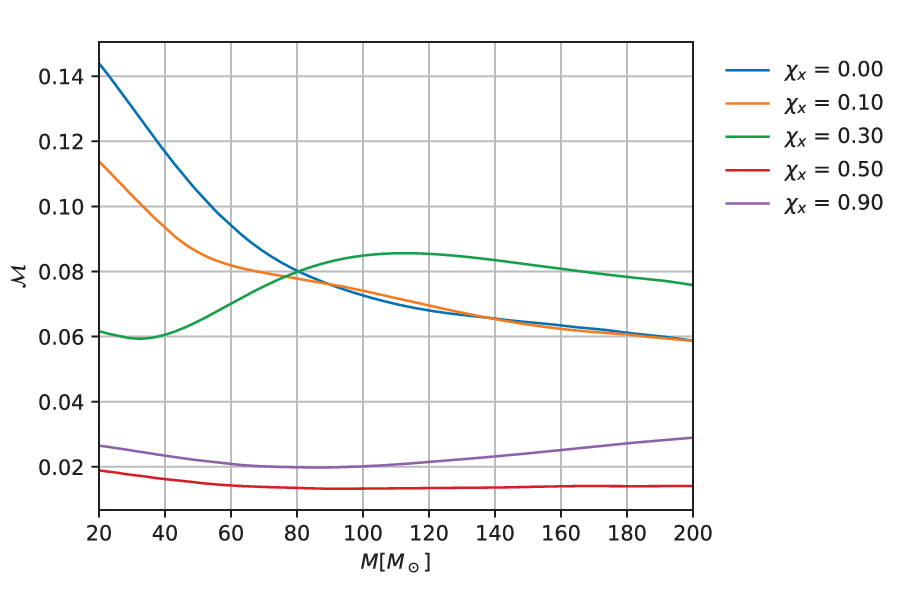}&
\includegraphics[width=0.5\textwidth]{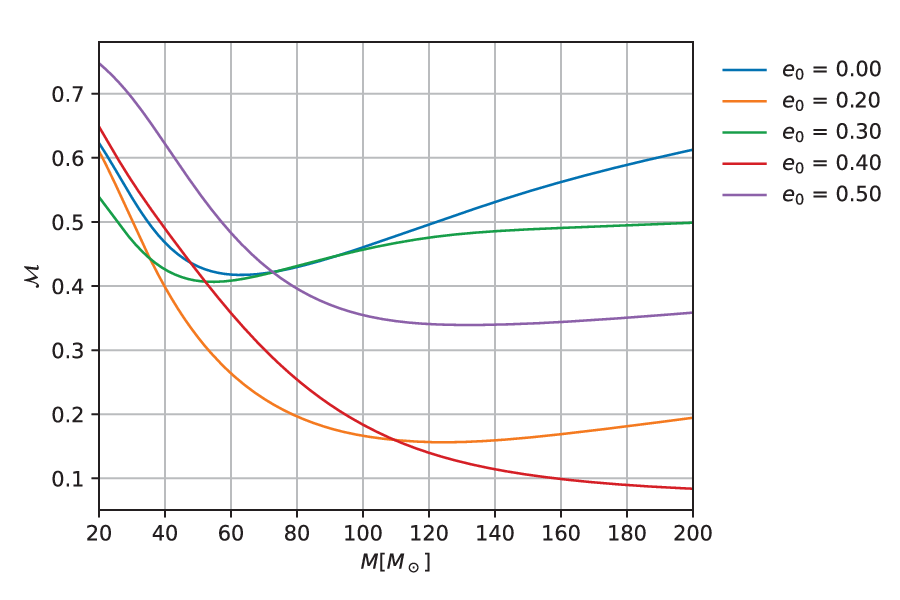}
\end{tabular}
\caption{Mismatch factors between the spin amplitude affected and orbit eccentricity affected waveforms and the referenced $\texttt{SEOBNRE}$ waveform modeling RIT:eBBH:1632. The left panel is for the spin amplitude affected waveforms corresponding to Fig.~\ref{fig9}. The right panel is for the orbit eccentricity affected waveforms corresponding to Fig.~\ref{fig10}.}
\label{fig11}
\end{figure*}

For LIGO-like ground-based detectors, the detected waveform is short. Only merger part is available. So it is some hard to distinguish spin precession and orbit eccentricity. Although we do not intend to investigate this issue in detail in the current work, we would like to mimic such detection through aligning waveforms at merger time and calculating corresponding mismatch factor. After time alignment, if the two waveforms admit different length, we just take the common time duration for the mismatch factor calculation. We plot the resulted mismatch factors between the spin amplitude affected (corresponding to Fig.~\ref{fig9}) and orbit eccentricity (corresponding to Fig.~\ref{fig10}) affected waveforms and the referenced $\texttt{SEOBNRE}$ waveform modeling RIT:eBBH:1632 in Fig.~\ref{fig11}. Just as we analyzed above, the orbit eccentricity results in much larger mismatch factor. So we expect our $\texttt{SEOBNRE}$ waveform model can help people to distinguish spin precession and orbit eccentricity in real data analysis. Off course, much more studies on the parameters degeneracy are needed.

\section{conclusion}
In this work, we apply the method proposed in \cite{PhysRevD.104.024046} to calculate the 2PN orbital dynamics of a general spinning precessing binary black hole system along eccentric orbit in EOB coordinates. We have also calculated the instantaneous part of the factorized waveform. Specifically, we have quantitatively analyzed the impact of non-perpendicular spin contributions on the waveforms.

We extend our previous constructed $\texttt{SEOBNRE}$ waveform model by incorporating the instantaneous part of general spin contributions and radiation reaction force correction. By comparing $\texttt{SEOBNRE}$ waveform model and $\texttt{SEOBNRv4PHM}$ waveform model with NR waveforms with the PSD of ground-based detectors, we found that the inclusion of non-perpendicular spin contributions led to a small improvement in waveform accuracy. This improvement mainly stems from the modification of the gravitational wave phase due to the non-perpendicular spin contributions, which first appears at the 1PN order but is not prominent. For ground-based detectors, since stellar-mass binary black hole systems are close to the merger phase when they enter the detection frequency band, the phase correction caused by non-perpendicular spin contributions is not significant, and the aligned spin waveform exhibits high accuracy at this stage. For future next-generation ground-based detectors and space-based detectors with lower detection frequency bands, the influence of this contribution may become important in waveform modeling.

As the first theoretical waveform model for generic spin-precession BBH with orbit eccentricity, we have also investigated the accuracy of the $\texttt{SEOBNRE}$ waveform model through comparing the model waveform to NR waveform. Both SXS \cite{SXSBBH} and RIT \cite{PhysRevD.105.124010} NR waveform catalogs are used in the current work. We have compared the new model with a NR waveform featuring significant spin precession and a relatively large initial eccentricity. We find that the new model approximately matches the numerical relativistic waveform, indicating that the corrections computed in this study can effectively describe this most general type of spin precession orbit with eccentricity. Good consistency between our model waveform and NR waveform is shown. In the near future our $\texttt{SEOBNRE}$ waveform model can help people as a tool to distinguish spin precession and orbit eccentricity in real data analysis.

In the following work we plan to improve our $\texttt{SEOBNRE}$ waveform model in several respects. Future efforts should involve constructing more comprehensive models, such as the 2PN complete EOB decomposed waveform presented in this paper and incorporating radiation-reaction forces in the dynamical system, followed by hyper-parameter tuning. Furthermore, calculating the contributions of the hereditary part of waveforms for general spinning precessing orbits is another topic worth discussing. Currently, the hereditary part of waveforms for generic equatorial orbits can be computed using quasi-Keplerian methods and methods based on small eccentricity expansions \cite{PhysRevD.100.044018}. However, for orbits with spin precession, constructing periodicity-based descriptions to estimate the contributions of their hereditary part requires further investigation.

\section*{Acknowledgments}
%|--------------------------------------------------------------------|
We would like to thank Alessandra Buonanno, Lijing Shao and Antoni Ramos Buades for many helpful discussions. This work was supported in part by the National Key Research and Development Program of China Grant
No. 2021YFC2203001, the Strategic Priority Research Program of the Chinese Academy of Sciences (No. XDB23000000) and in part by the NSFC (No.~11920101003 and No.~12021003). Z. Cao and Z.-H. Zhu were supported by ``the Interdiscipline Research Funds of Beijing Normal University".
%|--------------------------------------------------------------------|

\appendix
\begin{widetext}
\section{The radiation reaction force}\label{appA}
The radiation reaction force for general orbits up to 2PN order we get in this paper reads
\begin{align}
\pmb{F}= \left(f^r_{\text{ns}} + \frac{1}{c^3}f^r_{\text{so}} + \frac{1}{c^4}f^r_{\text{ss}}\right)\pmb{r} + \left( f^p_{\text{ns}} + \frac{1}{c^3}f^p_{\text{so}} + \frac{1}{c^4}f^p_{\text{ss}} \right)\pmb{p} + \frac{1}{c^3}f^L_{\text{so}}\pmb{L} + \frac{1}{c^4}f^1_{\text{ss}}\pmb{S}_1 + \frac{1}{c^4}f^2_{\text{ss}}\pmb{S}_2,
\end{align}
where the no-spin coefficients $f^r_{\text{ns}}, f^p_{\text{ns}}$ are
\begin{align}
%
% f^r_ns
%
&f^r_{\text{ns}} = -\frac{8 \nu ^2 p_r^3}{r^4}\left\{1 + \frac{1}{c^2}\left[-\frac{3}{14} (7 \nu -5) p^2-\frac{3508 \nu +1683}{420 r}\right] + \frac{1}{c^4}\left[\left(\frac{185 \nu ^2}{504}-\frac{2171 \nu }{504}+\frac{269}{504}\right) p^4 \right.\right.\nonumber \\
&\left.\left. + \left(\frac{19 \nu ^2}{9}+\frac{155 \nu
   }{36}-\frac{13}{6}\right) p^2 p_r^2+\left(\frac{2617 \nu ^2}{180}-\frac{5557 \nu }{72}-\frac{77837}{7560}\right)
   \frac{p^2}{r}+\left(-\frac{202 \nu ^2}{45}+\frac{3194 \nu }{315}+\frac{1151}{630}\right) \frac{p_r^2}{r} \right.\right.\nonumber \\
&\left.\left.+\left(\frac{281 \nu
   ^2}{54}-\frac{102167 \nu }{3780}+\frac{359}{1260}\right)\frac{1}{r^2}\right]\right\}, \\
%
% f^p_ns
%
&f^p_{\text{ns}}=-\frac{8\nu^2}{15r^3}\left\{10 p^2-39 p_r^2-\frac{22}{r} +\frac{1}{c^2}\left[\left(-\frac{3 \nu }{7}-\frac{279}{28}\right) p^4+\left(\frac{9 \nu }{7}+\frac{837}{28}\right) p^2 p_r^2+\left(-\frac{121 \nu
   }{14}-\frac{3833}{56}\right) \frac{p^2}{r} \right.\right.\nonumber \\
&\left.\left.+\left(\frac{45 \nu }{2}-\frac{225}{14}\right) p_r^4+\left(\frac{2445 \nu
   }{14}+\frac{9599}{28}\right) \frac{p_r^2}{r}+\frac{421 \nu }{14}+\frac{6213}{56}\frac{1}{r^2}\right] + \frac{1}{c^4}\left[\left(\frac{1319 \nu ^2}{42}+\frac{109609 \nu }{336}-\frac{45875}{1008}\right) p^6 \right.\right.\nonumber \\
&\left.\left.+\left(-\frac{1319 \nu ^2}{14}-\frac{109609 \nu
   }{112}+\frac{45875}{336}\right) p^4 p_r^2+\left(-\frac{2491 \nu ^2}{42}-\frac{100847 \nu }{336}+\frac{7595}{144}\right)
   \frac{p^4}{r}+\left(\frac{439 \nu ^2}{21}-\frac{2117 \nu }{12}+\frac{16775}{56}\right) \frac{p^2}{r^2} \right.\right.\nonumber \\
&\left.\left.+\left(-\frac{925 \nu
   ^2}{168}+\frac{10855 \nu }{168}-\frac{1345}{168}\right) p^2 p_r^4+\left(-\frac{15121 \nu ^2}{56}-\frac{63683 \nu
   }{42}+\frac{13037}{63}\right) \frac{p^2 p_r^2}{r}+\left(-\frac{401 \nu ^2}{42}-\frac{15405 \nu }{28}-\frac{44015}{56}\right)
   \frac{p_r^2}{r^2} \right.\right.\nonumber \\
&\left.\left.+\left(-\frac{95 \nu ^2}{3}-\frac{775 \nu }{12}+\frac{65}{2}\right) p_r^6+\left(-\frac{139 \nu ^2}{12}+\frac{129 \nu
   }{7}+\frac{1542}{7}\right) \frac{p_r^4}{r}+\left(\nu ^2+\frac{1216 \nu }{21}-\frac{47561}{378}\right)\frac{1}{r^3}\right] \right\}.
\end{align}
the spin-orbit coefficients $f^r_{\text{so}}, f^p_{\text{so}}, f^L_{\text{so}}$ are
\begin{align}
%
% f^r_so
%
&f^r_{\text{so}} = \frac{2p_r^3\nu}{15r^5L^2}\left[\pmb{L}\cdot\pmb{S}_1\left(-19 \delta +10 \nu +90 p^2 r (3 \delta -2 \nu -3)-105 r (3 \delta -2 \nu -3) p_r^2+19\right) \right. \nonumber \\
&\qquad\left. + \pmb{L}\cdot\pmb{S}_2\left(19 \delta +10 \nu -90 p^2 r (3 \delta +2 \nu +3)+105 r (3 \delta +2 \nu +3) p_r^2+19\right)\right], \\
%
% f^p_so
%
&f^p_{\text{so}} = \frac{2\nu}{15r^4L^2}\left\{\pmb{L}\cdot\pmb{S}_1\left[2 p^4 r (\delta +11 \nu -1)+p^2 \left(2 (53 \delta -73 \nu -53)+3 r (41 \delta -74 \nu -41) p_r^2\right) \right.\right.\nonumber \\
&\qquad\left.\left.+p_r^2 \left(-125 \delta +156 \nu
   +10 r (-17 \delta +23 \nu +17) p_r^2+125\right]\right) \right. \nonumber \\
&\qquad\left. + \pmb{L}\cdot\pmb{S}_2\left[-2 p^4 r (\delta -11 \nu +1)-p^2 \left(2 (53 (\delta +1)+73 \nu )+3 r (41 (\delta +1)+74 \nu ) p_r^2\right) \right.\right.\nonumber \\
&\qquad\left.\left.+p_r^2 \left(125 (\delta +1)+156
   \nu +10 r (17 (\delta +1)+23 \nu ) p_r^2\right)\right]\right\}, \\
%
% f^L_so
%
&f^L_{\text{so}} = \frac{2p_r^2}{15r^4L^2}\left\{ \frac{p_r\nu}{r}\pmb{r}\cdot\pmb{S}_1\left[-19 \delta +10 \nu +15 p^2 r (-3 \delta +2 \nu +3)+19\right] + \frac{p_r\nu}{r}\pmb{r}\cdot\pmb{S}_2\left[19 (\delta +1)+10 \nu +15 p^2 r (3 \delta +2 \nu +3)\right] \right.\nonumber \\
&\qquad\left.+\pmb{p}\cdot\pmb{S}_1\left[\nu  (19 \delta -10 \nu -19)-15 p^2 r \left(5 (\delta -3) \nu -3 \delta -2 \nu ^2+3\right)+15 r \left(8 \delta  \nu -3 \delta -4 \nu ^2-18
   \nu +3\right) p_r^2\right] \right.\nonumber \\
&\qquad\left.+\pmb{p}\cdot\pmb{S}_2\left[-\nu  (19 \delta +10 \nu +19)+15 p^2 r \left(5 (\delta +3) \nu -3 (\delta +1)+2 \nu ^2\right)-15 r \left(8 \delta  \nu -3 \delta +4 \nu
   ^2+18 \nu -3\right) p_r^2\right]\right\}
\end{align}
the spin-spin coefficients $f^r_{\text{ss}}, f^p_{\text{ss}}, f^1_{\text{ss}}, f^2_{\text{ss}}$ are
\begin{align}
%
% f^r_ss
%
&f^r_{\text{ss}}=\frac{2p_r^2}{15r^7}\left[15 r p_r \left(\pmb{S}_1\cdot\pmb{S}_1 \left(7 \delta +2 \nu ^2+14 \nu -7\right)+4 (\nu -7) \nu  \pmb{S}_1\cdot\pmb{S}_2+\pmb{S}_2\cdot\pmb{S}_2 \left(-7 \delta +2 \nu ^2+14 \nu -7\right)\right) \right.\nonumber \\
&\qquad\left.+\nu  (244 \nu  \pmb{p}\cdot\pmb{S}_1 \pmb{r}\cdot\pmb{S}_1+488 \nu  \pmb{p}\cdot\pmb{S}_1 \pmb{r}\cdot\pmb{S}_2-105 \pmb{p}\cdot\pmb{S}_1 \pmb{r}\cdot\pmb{S}_2+105 \pmb{p}\cdot\pmb{S}_2 \pmb{r}\cdot\pmb{S}_1+244 \nu  \pmb{p}\cdot\pmb{S}_2 \pmb{r}\cdot\pmb{S}_2)\right], \\
%
% f^p_ss
%
&f^p_{\text{ss}}=-\frac{1}{30r^8}\left[-171 \delta  p^2 r^3 \pmb{S}_1\cdot\pmb{S}_1-400 \nu ^2 p^2 r^3 \pmb{S}_1\cdot\pmb{S}_1-342 \nu  p^2 r^3 \pmb{S}_1\cdot\pmb{S}_1+171 p^2 r^3 \pmb{S}_1\cdot\pmb{S}_1-800 \nu ^2 p^2 r^3 \pmb{S}_1\cdot\pmb{S}_2 \right.\nonumber\\
&\qquad\left.+756 \nu  p^2 r^3 \pmb{S}_1\cdot\pmb{S}_2+171 \delta  p^2 r^3 \pmb{S}_2\cdot\pmb{S}_2-400 \nu ^2 p^2 r^3 \pmb{S}_2\cdot\pmb{S}_2-342 \nu  p^2 r^3 \pmb{S}_2\cdot\pmb{S}_2+171 p^2 r^3 \pmb{S}_2\cdot\pmb{S}_2\right.\nonumber\\
&\qquad\left.+1035 \delta  r^3 \pmb{S}_1\cdot\pmb{S}_1 p_r^2+632 \nu ^2 r^3 \pmb{S}_1\cdot\pmb{S}_1 p_r^2+2070 \nu  r^3 \pmb{S}_1\cdot\pmb{S}_1 p_r^2-1035 r^3 \pmb{S}_1\cdot\pmb{S}_1 p_r^2+1264 \nu ^2 r^3 \pmb{S}_1\cdot\pmb{S}_2 p_r^2\right.\nonumber\\
&\qquad\left.-4500 \nu  r^3 \pmb{S}_1\cdot\pmb{S}_2 p_r^2-1035 \delta  r^3 \pmb{S}_2\cdot\pmb{S}_2 p_r^2+632 \nu ^2 r^3 \pmb{S}_2\cdot\pmb{S}_2 p_r^2+2070 \nu  r^3 \pmb{S}_2\cdot\pmb{S}_2 p_r^2-1035 r^3 \pmb{S}_2\cdot\pmb{S}_2 p_r^2\right.\nonumber\\
&\qquad\left.+480 \nu ^2 r (\pmb{r}\cdot\pmb{S}_1)^2 p_r^2+960 \nu ^2 r (\pmb{r}\cdot\pmb{S}_1) (\pmb{r}\cdot\pmb{S}_2) p_r^2+480 \nu ^2 r (\pmb{r}\cdot\pmb{S}_2)^2 p_r^2+48 (\pmb{p}\cdot\pmb{S}_1)^2 r^3 \left(9 \delta +14 \nu ^2+3 \nu -9\right)\right.\nonumber\\
&\qquad\left.+96 \nu  (14 \nu -3) (\pmb{p}\cdot\pmb{S}_1) (\pmb{p}\cdot\pmb{S}_2) r^3-48 (\pmb{p}\cdot\pmb{S}_2)^2 r^3 \left(9 \delta -14 \nu ^2-3 \nu +9\right)+585 \delta  r^2 \pmb{S}_1\cdot\pmb{S}_1+704 \nu ^2 r^2 \pmb{S}_1\cdot\pmb{S}_1\right.\nonumber\\
&\qquad\left.+1170 \nu  r^2 \pmb{S}_1\cdot\pmb{S}_1-585 r^2 \pmb{S}_1\cdot\pmb{S}_1+1408 \nu ^2 r^2 \pmb{S}_1\cdot\pmb{S}_2-2364 \nu  r^2 \pmb{S}_1\cdot\pmb{S}_2-585 \delta  r^2 \pmb{S}_2\cdot\pmb{S}_2+704 \nu ^2 r^2 \pmb{S}_2\cdot\pmb{S}_2\right.\nonumber\\
&\qquad\left.+1170 \nu  r^2 \pmb{S}_2\cdot\pmb{S}_2-585 r^2 \pmb{S}_2\cdot\pmb{S}_2-1368 \delta  (\pmb{r}\cdot\pmb{S}_1)^2-624 \nu ^2 (\pmb{r}\cdot\pmb{S}_1)^2-2736 \nu  (\pmb{r}\cdot\pmb{S}_1)^2+1368 (\pmb{r}\cdot\pmb{S}_1)^2\right.\nonumber\\
&\qquad\left.-1248 \nu ^2 (\pmb{r}\cdot\pmb{S}_1) (\pmb{r}\cdot\pmb{S}_2)+5472 \nu  (\pmb{r}\cdot\pmb{S}_1) (\pmb{r}\cdot\pmb{S}_2)+1368 \delta  (\pmb{r}\cdot\pmb{S}_2)^2-624 \nu ^2 (\pmb{r}\cdot\pmb{S}_2)^2-2736 \nu  (\pmb{r}\cdot\pmb{S}_2)^2+1368 (\pmb{r}\cdot\pmb{S}_2)^2\right], \\
%
% f^1_ss
%
&f^1_{\text{ss}}=\frac{2p_r^2\nu}{15r^6}\left[244 \nu  \pmb{r}\cdot\pmb{S}_1 p_r+105 \pmb{r}\cdot\pmb{S}_2 p_r-244 \
\nu  \pmb{p}\cdot\pmb{S}_1 r-105 \pmb{p}\cdot\pmb{S}_2 r\right], \\
%
% f^2_ss
%
&f^2_{\text{ss}}=\frac{2p_r^2\nu}{15r^6}\left[p_r (488 \nu  \pmb{r}\cdot\pmb{S}_1-105 \pmb{r}\cdot\pmb{S}_1+244 \nu  \pmb{r}\cdot\pmb{S}_2)+(105-488 \nu ) \pmb{p}\cdot\pmb{S}_1 r-244 \nu  \pmb{p}\cdot\pmb{S}_2 r\right].
\end{align}

\section{The quasi-circular waveform correction}\label{appB}
The quasi-circular corrections resulted by the spin miss-aligned part are given by
\begin{align}
%
% \rho_{22}
%
&\rho_{22}= 1+\frac{1}{84} v_\Omega^2 (21 \chi_{a\lambda}+55 \nu +21 i \chi_{an}+21 \delta  (\chi_{s\lambda}+i \chi_{sn})-86)-\frac{2}{3} v_\Omega^3 (\delta  \chi_{ae}-\nu\chi_{se} +\chi_{se}) \nonumber \\
&+v_\Omega^4 \left(\delta  \chi_{a\lambda} \chi_{s\lambda} \nu +\frac{7 \delta  \chi_{a\lambda} \chi_{s\lambda}}{16}+\frac{11 \delta  \chi_{s\lambda} \nu }{336}+\frac{13 \delta  \chi_{s\lambda}}{28}-2 \chi_{ae}^2 \nu +\frac{\chi_{ae}^2}{2}+\delta  \chi_{ae} \chi_{se}+\frac{\chi_{se}^2}{2}-\chi_{a\lambda}^2 \nu +\frac{7 \chi_{a\lambda}^2}{32} \right.\nonumber \\
&\left.-\frac{121 \chi_{a\lambda} \nu }{336}+\frac{13 \chi_{a\lambda}}{28}-\chi_{s\lambda}^2 \nu ^2+\frac{9 \chi_{s\lambda}^2 \nu }{8}+\frac{7 \chi_{s\lambda}^2}{32}+\frac{19583 \nu ^2}{42336}-\frac{33025 \nu }{21168}+\frac{\chi_{an}^2}{32}-\frac{25}{16} i \delta  \chi_{s\lambda} \chi_{an}+6 i \chi_{a\lambda} \nu  \chi_{an} \right.\nonumber \\
&\left.-\frac{25 i \chi_{a\lambda} \chi_{an}}{16}-\frac{317 i \nu  \chi_{an}}{112}-\delta  \nu  \chi_{an} \chi_{sn}+\frac{\delta  \chi_{an} \chi_{sn}}{16}+\frac{37 i \chi_{an}}{42}+\nu ^2 \chi_{sn}^2-\frac{9 \nu  \chi_{sn}^2}{8}+\frac{\chi_{sn}^2}{32}-\frac{25}{16} i \delta  \chi_{a\lambda} \chi_{sn}\right.\nonumber\\
&\left.-\frac{167}{336} i \delta  \nu  \chi_{sn}+\frac{37 i \delta  \chi_{sn}}{42}+\frac{1}{4} i \chi_{s\lambda} \nu  \chi_{sn}-\frac{25 i \chi_{s\lambda} \chi_{sn}}{16}-\frac{20555}{10584}\right), \\
%
% f^s_{21}
%
&f^s_{21}=v_\Omega\left(-\frac{3 \chi_{ae}}{2 \delta }-\frac{3 \chi_{se}}{2}\right) + v_\Omega^2\left(\frac{\chi_{s\lambda} \nu }{\delta }-6 i \chi_{an}+\frac{2 i \nu  \chi_{sn}}{\delta }-\frac{6 i \chi_{sn}}{\delta }\right) + v_\Omega^3\left(-\frac{6 \chi_{ae} \chi_{a\lambda} \nu }{\delta }+\frac{3 \chi_{ae} \chi_{a\lambda}}{2 \delta } \right.\nonumber \\
&\left.+\frac{131 \chi_{ae} \nu }{84 \delta }+\frac{61 \chi_{ae}}{12 \delta }-3 \chi_{ae} \chi_{s\lambda} \nu +\frac{3 \chi_{ae} \chi_{s\lambda}}{2}-\frac{30 i \chi_{ae} \nu  \chi_{an}}{\delta }+\frac{15 i \chi_{ae} \chi_{an}}{2 \delta }+3 i \chi_{ae} \nu  \chi_{sn}+\frac{15 i \chi_{ae} \chi_{sn}}{2} \right.\nonumber\\
&\left.+\frac{6 \chi_{se} \chi_{s\lambda} \nu ^2}{\delta }-\frac{6 \chi_{se} \chi_{s\lambda} \nu }{\delta }+\frac{3 \chi_{se} \chi_{s\lambda}}{2 \delta }-3 \chi_{se} \chi_{a\lambda} \nu +\frac{3 \chi_{se} \chi_{a\lambda}}{2}+\frac{79 \chi_{se} \nu }{84}+3 i \chi_{se} \nu  \chi_{an}+\frac{15 i \chi_{se} \chi_{an}}{2}\right.\nonumber\\
&\left.-\frac{6 i \chi_{se} \nu ^2 \chi_{sn}}{\delta }+\frac{6 i \chi_{se} \nu  \chi_{sn}}{\delta }+\frac{15 i \chi_{se} \chi_{sn}}{2 \delta }+\frac{61 \chi_{se}}{12}\right)\\
%
% f^s_{33}
%
&f^s_{33}=v_\Omega^2\left(-\frac{16 \chi_{s\lambda} \nu }{9 \delta }-\frac{16 i \nu  \chi_{sn}}{9 \delta }\right)+v_\Omega^3\left(\frac{19 \chi_{ae} \nu }{2 \delta }-\frac{2 \chi_{ae}}{\delta }+\frac{5 \chi_{se} \nu }{2}-2 \chi_{se}\right) \\
%
% \rho_{32}
%
&\rho_{32}= 1+\frac{4v_\Omega \chi_{se} \nu }{3-9 \nu }+v_\Omega^2\left[-\frac{7 \delta  \chi_{s\lambda} \nu ^2}{8 (1-3 \nu )^2}-\frac{17 \delta  \chi_{s\lambda} \nu }{24 (1-3 \nu )^2}+\frac{\delta  \chi_{s\lambda}}{3 (1-3 \nu )^2}-\frac{16 \chi_{se}^2 \nu ^2}{9 (1-3 \nu )^2}+\frac{17 \chi_{a\lambda} \nu ^2}{8 (1-3 \nu )^2}\right.\nonumber\\
&\left.-\frac{41 \chi_{a\lambda} \nu }{24 (1-3 \nu )^2}+\frac{\chi_{a\lambda}}{3 (1-3 \nu )^2}-\frac{45 i \nu ^2 \chi_{an}}{2 (1-3 \nu )^2}+\frac{27 i \nu  \chi_{an}}{2 (1-3 \nu )^2}-\frac{2 i \chi_{an}}{(1-3 \nu )^2}-\frac{i \delta  \nu ^2 \chi_{sn}}{(1-3 \nu )^2}+\frac{19 i \delta  \nu  \chi_{sn}}{3 (1-3 \nu )^2}-\frac{2 i \delta  \chi_{sn}}{(1-3 \nu )^2}\right]\\
%
% \rho_{31}
%
&f^s_{31}=v_\Omega^2\left(\frac{16 \chi_{s\lambda} \nu }{\delta }-\frac{16 i \nu  \chi_{sn}}{\delta }\right)+v_\Omega^3\left(\frac{11 \chi_{ae} \nu }{2 \delta }-\frac{2 \chi_{ae}}{\delta }+\frac{13 \chi_{se} \nu }{2}-2 \chi_{se}\right) \\
%
% \rho_{44}
%
&\rho_{44}=1+v_\Omega^2\left[\frac{81 \delta  \chi_{s\lambda} \nu }{256 (3 \nu -1)}-\frac{81 \chi_{a\lambda} \nu }{256 (3 \nu -1)}+\frac{175 \nu ^2}{88 (3 \nu -1)}-\frac{587 \nu }{132 (3 \nu -1)}+\frac{269}{220 (3 \nu -1)}-\frac{81 i \nu  \chi_{an}}{256 (3 \nu -1)}+\frac{81 i \delta  \nu  \chi_{sn}}{256 (3 \nu -1)}\right] \\
%
% f^s_{43}
%
&f^s_{43}=\frac{5v_\Omega\eta}{2(1-2\eta)}\left(\chi_{se}-\frac{\chi_{ae}}{\delta}\right)\\
%
% \rho_{42}
%
&\rho_{42}=1+v_\Omega^2\left[-\frac{39 \delta  \chi_{s\lambda} \nu }{32 (3 \nu -1)}+\frac{39 \chi_{a\lambda} \nu }{32 (3 \nu -1)}+\frac{19 \nu ^2}{88 (3 \nu -1)}-\frac{353 \nu }{132 (3 \nu -1)}+\frac{191}{220 (3 \nu -1)}-\frac{21 i \nu  \chi_{an}}{16 (3 \nu -1)}+\frac{21 i \delta  \nu  \chi_{sn}}{16 (3 \nu -1)}\right] \\
%
% f_{41}
%
&f^s_{41}=\frac{5v_\Omega\eta}{2(1-2\eta)}\left(\chi_{se}-\frac{\chi_{ae}}{\delta}\right),
\end{align}

\section{test the spin-precession $\texttt{SEOBNRE}$ waveform model against the spin-aligned BBH}\label{appnew}
In \cite{Liu_2022} we have developed higher modes for $\texttt{SEOBNRE}$ waveform model, including (2,1), (3,3) and (4,4). In the current paper we construct (5,5) mode also. Although the $\texttt{SEOBNRE}$ waveform model proposed here is for generic eccentric spin-precession BBHs, it also works for spin-aligned BBHs. As an example we show the comparison between the NR waveform SXS:BBH:1374 and the corresponding $\texttt{SEOBNRE}$ waveform in Fig.~\ref{fig12}.
\begin{figure*}[t]
\centering
\begin{tabular}{c}
\includegraphics[width=\textwidth]{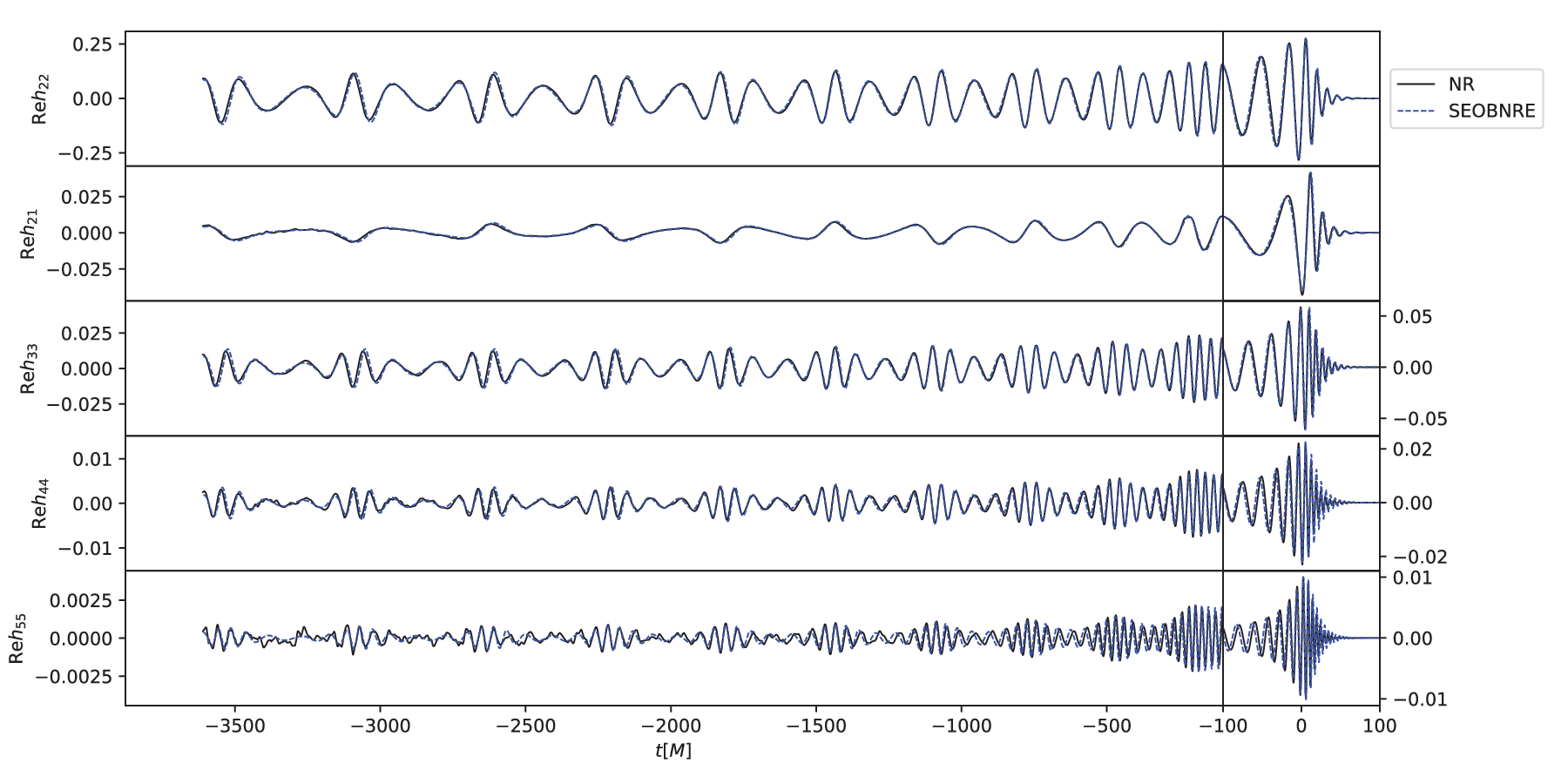}
\end{tabular}
\caption{Waveforms comparison between the NR waveform SXS:BBH:1374 and the corresponding $\texttt{SEOBNRE}$ waveform. This is a spinless BBH with mass ratio $q = 3$.}
\label{fig12}
\end{figure*}
\section{SXS waveforms used in the current work}\label{appC}
\begin{longtable}{ccccccc}
    \caption{The NR waveforms used in this work. Here, we list the parameters for each waveform, including mass ratio $q$, initial frequency $Mf_{\text{ini}}$, and initial spin configuration. Additionally, we provide the maximum mismatch between the two EOB models and these waveforms. The subscription `old' means $\texttt{SEOBNRv4PHM}$ waveform model, and `new' means $\texttt{SEOBNRE}$ waveform model.}\label{table_appC} \\
    \hline
    \hline
    SXS ID & $q$ & $\pmb{\chi}_1$ & $\pmb{\chi}_2$ & $10^3Mf_{\text{ini}}$ & $\max\mathcal{M}_{\text{old}}$ & $\max\mathcal{M}_{\text{new}}$ \\
    \hline
    \endfirsthead
    \multicolumn{7}{c}{\bfseries\small \tablename\ \thetable\ {continue}}\\
    \hline
    \hline
    SXS iD & $q$ & $\pmb{\chi}_1$ & $\pmb{\chi}_2$ & $10^3Mf_{\text{ini}}$ & $1-\text{unf}_{\text{old}}$ & $1-\text{unf}_{\text{new}}$ \\
    \hline
    \endhead
    \hline
    \multicolumn{7}{r@{}}{\textit{to next page}}
    \endfoot
    \hline
    \hline
    \endlastfoot    %
    \input{sxsprectable.tex}
\end{longtable}

\end{widetext}
\bibliography{refs}
\end{document}

%% file: sxsprectable.tex
0401 & 1.50 & (0.38,0.42,-0.57) & (-0.03,0.03,-0.80) & 0.45 & 0.16\% & 0.19\% \\
0622 & 1.20 & (0.42,0.42,0.60) & (0.54,0.54,0.37) & 0.53 & 2.17\% & 2.15\% \\
0623 & 1.11 & (0.89,0.00,0.11) & (0.89,0.00,-0.12) & 0.31 & 2.48\% & 2.31\% \\
%0624 & 1.20 & (-0.42,-0.42,0.60) & (0.31,0.31,0.73) & 0.53 & 37.42\% & 30.79\% \\
0628 & 2.06 & (-0.38,0.75,-0.10) & (0.08,-0.40,-0.30) & 0.61 & 0.60\% & 0.71\% \\
0632 & 1.08 & (0.72,0.24,-0.09) & (0.63,-0.49,-0.39) & 0.51 & 1.21\% & 1.42\% \\
0633 & 1.34 & (0.38,-0.16,-0.66) & (-0.79,-0.06,0.26) & 0.50 & 3.80\% & 1.44\% \\
0651 & 1.33 & (-0.00,0.00,-0.00) & (0.58,0.55,0.00) & 0.48 & 0.27\% & 0.25\% \\
%0656 & 1.33 & (0.43,-0.68,0.01) & (-0.07,-0.03,0.80) & 0.49 & 13.78\% & 9.80\% \\
%0705 & 2.00 & (-0.03,-0.03,0.80) & (0.57,-0.56,0.01) & 0.52 & 19.28\% & 3.82\% \\
%0711 & 2.00 & (0.48,-0.64,-0.03) & (0.49,0.63,0.08) & 0.50 & 76.59\% & 1.02\% \\
0715 & 2.00 & (-0.79,-0.10,-0.04) & (0.30,-0.74,0.08) & 0.50 & 0.92\% & 0.85\% \\
0721 & 2.00 & (0.31,0.74,-0.03) & (-0.79,0.11,0.08) & 0.49 & 1.23\% & 1.00\% \\
0753 & 1.00 & (0.44,-0.35,0.57) & (-0.00,-0.00,-0.00) & 0.49 & 0.20\% & 0.18\% \\
0787 & 2.00 & (-0.48,-0.25,-0.59) & (0.09,-0.79,0.06) & 0.48 & 2.49\% & 1.26\% \\
0888 & 2.00 & (-0.77,-0.21,-0.01) & (-0.50,-0.21,0.59) & 0.49 & 2.51\% & 1.78\% \\
0891 & 1.00 & (0.01,-0.04,-0.80) & (-0.56,-0.07,0.56) & 0.46 & 0.62\% & 0.45\% \\
0904 & 2.00 & (-0.00,-0.03,0.80) & (0.56,-0.08,0.57) & 0.52 & 0.27\% & 0.24\% \\
0906 & 2.00 & (0.57,-0.56,0.01) & (0.42,-0.33,0.59) & 0.50 & 1.22\% & 1.26\% \\
0940 & 2.00 & (0.00,-0.00,-0.00) & (-0.54,-0.16,-0.57) & 0.49 & 0.70\% & 0.66\% \\
0951 & 2.00 & (0.38,0.70,-0.01) & (0.07,0.59,-0.54) & 0.49 & 6.16\% & 2.11\% \\
0961 & 2.00 & (0.00,0.00,-0.00) & (-0.28,0.75,0.00) & 0.50 & 0.45\% & 0.41\% \\
1106 & 1.68 & (0.72,-0.00,0.36) & (-0.14,-0.07,0.05) & 0.29 & 6.14\% & 0.92\% \\
1184 & 1.00 & (0.06,-0.04,0.85) & (0.40,0.70,0.01) & 0.49 & 1.72\% & 1.48\% \\
1185 & 1.00 & (-0.73,0.43,-0.00) & (-0.03,-0.05,-0.80) & 0.45 & 0.35\% & 0.33\% \\
1188 & 1.00 & (0.62,0.58,0.05) & (0.19,-0.78,-0.05) & 0.48 & 0.10\% & 0.13\% \\
1189 & 1.00 & (0.61,0.59,-0.05) & (-0.77,0.23,0.06) & 0.47 & 0.31\% & 0.23\% \\
1190 & 1.00 & (0.74,0.42,-0.00) & (-0.03,0.05,-0.80) & 0.45 & 0.21\% & 0.31\% \\
1191 & 1.00 & (0.20,-0.82,0.06) & (-0.77,0.23,-0.06) & 0.48 & 0.35\% & 0.41\% \\
1192 & 1.00 & (-0.01,-0.85,0.00) & (0.06,-0.00,-0.80) & 0.46 & 0.28\% & 0.36\% \\
1193 & 2.00 & (0.54,-0.66,0.03) & (0.35,-0.78,0.00) & 0.50 & 2.54\% & 2.86\% \\
1194 & 2.00 & (0.51,-0.68,0.05) & (-0.84,0.12,-0.08) & 0.49 & 1.30\% & 1.14\% \\
1195 & 1.00 & (-0.81,0.24,-0.06) & (0.20,-0.82,0.06) & 0.48 & 0.59\% & 0.52\% \\
1196 & 1.00 & (0.23,-0.82,0.01) & (0.23,-0.82,0.01) & 0.49 & 2.08\% & 1.84\% \\
1197 & 2.00 & (-0.84,-0.11,-0.04) & (0.32,-0.78,0.09) & 0.50 & 1.25\% & 1.13\% \\
1198 & 1.00 & (0.43,-0.73,0.01) & (-0.07,-0.03,0.85) & 0.49 & 1.68\% & 1.49\% \\
1199 & 2.00 & (0.39,-0.75,0.01) & (0.09,0.03,-0.84) & 0.48 & 1.87\% & 1.67\% \\
1200 & 2.00 & (-0.84,-0.14,0.01) & (-0.85,0.09,-0.01) & 0.48 & 2.02\% & 2.68\% \\
1201 & 1.00 & (-0.03,-0.05,-0.85) & (-0.73,0.44,-0.00) & 0.45 & 0.25\% & 0.35\% \\
1202 & 2.00 & (-0.01,-0.04,-0.85) & (-0.73,0.44,0.00) & 0.46 & 0.06\% & 0.08\% \\
1203 & 2.00 & (-0.85,0.04,-0.00) & (-0.02,-0.09,-0.85) & 0.47 & 2.51\% & 2.31\% \\
1204 & 2.00 & (0.33,0.78,0.04) & (0.32,-0.78,-0.08) & 0.50 & 1.33\% & 0.88\% \\
1205 & 1.00 & (-0.81,0.24,0.06) & (0.61,0.59,-0.06) & 0.47 & 0.35\% & 0.24\% \\
1206 & 1.00 & (0.20,-0.82,-0.05) & (0.62,0.58,0.06) & 0.48 & 0.32\% & 0.20\% \\
%1207 & 2.00 & (0.52,-0.67,-0.04) & (0.52,0.67,0.09) & 0.50 & 76.84\% & 1.45\% \\
1208 & 1.00 & (-0.03,0.05,-0.85) & (0.75,0.41,-0.00) & 0.45 & 0.51\% & 0.38\% \\
1209 & 2.00 & (0.06,-0.01,0.85) & (0.18,0.83,0.01) & 0.53 & 1.20\% & 1.37\% \\
%1210 & 1.00 & (0.42,0.74,0.01) & (0.06,-0.04,0.85) & 0.49 & 18.26\% & 2.36\% \\
1211 & 2.00 & (-0.03,0.03,-0.85) & (0.74,0.41,-0.00) & 0.46 & 0.26\% & 0.13\% \\
1212 & 2.00 & (0.17,0.83,0.01) & (0.10,-0.04,0.84) & 0.50 & 2.49\% & 2.24\% \\
1213 & 2.00 & (0.46,0.72,0.00) & (-0.07,0.06,-0.84) & 0.48 & 1.57\% & 1.49\% \\
1214 & 1.00 & (0.60,0.60,0.00) & (0.60,0.60,0.00) & 0.48 & 2.89\% & 1.97\% \\
1215 & 2.00 & (0.30,0.79,0.02) & (0.50,0.69,-0.01) & 0.51 & 1.52\% & 1.59\% \\
1216 & 2.00 & (0.40,-0.75,0.01) & (0.09,0.03,-0.79) & 0.48 & 1.97\% & 1.78\% \\
1217 & 1.00 & (0.23,-0.82,0.01) & (0.21,-0.77,0.01) & 0.49 & 1.53\% & 1.79\% \\
1218 & 1.00 & (0.20,-0.82,-0.05) & (0.58,0.55,0.05) & 0.48 & 0.28\% & 0.19\% \\
1219 & 1.00 & (0.42,-0.74,0.01) & (-0.06,-0.02,0.80) & 0.49 & 1.58\% & 1.34\% \\
1909 & 4.00 & (0.00,-0.00,-0.00) & (-0.32,-0.73,0.01) & 0.56 & 0.07\% & 0.07\% \\
1918 & 4.00 & (-0.80,-0.07,0.02) & (-0.05,-0.10,-0.79) & 0.54 & 2.25\% & 1.40\% \\
%1923 & 4.00 & (-0.76,0.23,0.07) & (-0.34,0.73,-0.02) & 0.56 & 5.88\% & 5.61\% \\
%1926 & 4.00 & (0.79,0.13,0.03) & (-0.05,-0.11,0.79) & 0.55 & 2.21\% & 4.59\% \\
1930 & 4.00 & (0.34,0.72,0.00) & (-0.09,0.09,-0.79) & 0.54 & 1.86\% & 1.32\% \\
1968 & 4.00 & (-0.34,0.43,0.58) & (0.11,0.03,0.79) & 0.58 & 1.34\% & 1.08\% \\
1969 & 4.00 & (0.15,-0.53,0.58) & (0.11,-0.00,-0.79) & 0.57 & 1.01\% & 1.11\% \\
1971 & 4.00 & (-0.19,-0.53,0.57) & (-0.08,0.05,0.79) & 0.58 & 0.63\% & 0.71\% \\
1981 & 4.00 & (0.08,-0.59,-0.53) & (-0.79,-0.06,-0.08) & 0.52 & 3.93\% & 3.81\% \\
%1989 & 4.00 & (-0.54,-0.17,-0.56) & (-0.01,-0.07,-0.80) & 0.51 & 2.66\% & 6.46\% \\
1999 & 4.00 & (0.20,-0.78,0.00) & (0.69,-0.38,0.14) & 0.55 & 2.11\% & 2.42\% \\
2009 & 4.00 & (0.29,0.74,0.01) & (0.09,-0.08,0.79) & 0.55 & 2.07\% & 1.67\% \\
2012 & 4.00 & (-0.45,0.66,0.04) & (-0.14,-0.01,-0.79) & 0.54 & 2.70\% & 2.15\% \\
2060 & 4.00 & (0.00,-0.00,-0.00) & (-0.23,-0.52,-0.56) & 0.55 & 0.30\% & 0.31\% \\
2080 & 4.00 & (0.78,-0.19,0.05) & (-0.79,0.06,-0.08) & 0.54 & 1.39\% & 1.40\%